\documentclass[nofootinbib,eqsecnum,superscriptaddress,amsmath,amssymb,aps,prd,11pt]{revtex4-2}

\usepackage{amsfonts}
\usepackage{graphicx}

\usepackage{bbm} 
\usepackage{color} 

\usepackage{footnotebackref} 

\DeclareMathOperator{\Tr}{Tr} 

\newcommand{\abs}[1]{{\left|{#1}\right|}} 
\newcommand{\inner}[2]{{\langle {#1}\vert {#2} \rangle}} 
\newcommand{\ket}[1]{\vert{#1}\rangle} 
\newcommand{\bra}[1]{\langle{#1}\vert} 

\newcommand{\Path}[1]{{\mathrm{Path}_{#1}}} 

\newcommand{\secref}[1]{Sec.~\ref{#1}}

\newcommand{\eqnref}[1]{(\ref{#1})}
\newcommand{\figref}[1]{Fig.~\ref{#1}}

\newcommand{\ftref}[1]{Footnote~\ref{#1}} 

\begin{document}
\count\footins = 1000 

\title{Delayed-choice quantum erasers and the Einstein-Podolsky-Rosen paradox}


\author{Dah-Wei Chiou}
\email{dwchiou@gmail.com}
\affiliation{Department of Physics, National Sun Yat-sen University, Kaohsiung 80424, Taiwan}
\affiliation{Center for Condensed Matter Sciences, National Taiwan University, Taipei 10617, Taiwan}


\begin{abstract}
Considering the delayed-choice quantum eraser using a Mach-Zehnder interferometer with a nonsymmetric beam splitter, we explicitly demonstrate that it shares exactly the same formal structure with the Einstein-Podolsky-Rosen-Bohm (EPR-Bohm) experiment. Therefore, the effect of quantum erasure can be understood in terms of the standard EPR correlation. Nevertheless, the quantum eraser still raises a conceptual issue beyond the standard EPR paradox, if counterfactual reasoning is taken into account. Furthermore, the quantum eraser experiments can be classified into two major categories: the \emph{entanglement quantum eraser} and the \emph{Scully-Dr{\"u}hl-type quantum eraser}. These two types are formally equivalent to each other, but conceptually the latter presents a ``mystery'' more prominent than the former. In the Scully-Dr{\"u}hl-type quantum eraser, the statement that the which-way information can be influenced by the delayed-choice measurement is not purely a consequence of counterfactual reasoning but bears some factual significance. Accordingly, it makes good sense to say that the ``record'' of the which-way information is ``erased'' if the potentiality to yield a conclusive outcome that discriminates the record is eliminated by the delayed-choice measurement. We also reconsider the quantum eraser in the many-worlds interpretation (MWI), making clear the conceptual merits and demerits of the MWI.
\end{abstract}

\maketitle



\section{Introduction}
The idea of the delayed-choice quantum eraser was first proposed by Scully and Dr{\"u}hl in 1982 \cite{PhysRevA.25.2208}. The quantum eraser is an interferometer experiment in which the which-way information of each single quanton (i.e., quantum object such as photon) is ``marked'' and therefore the interference fringe pattern is not seen, but the which-way information can later be ``erased'' and correspondingly the interference pattern can be ``recovered'', apparently exhibiting some kind of retrocausality.
The first experiment of the delayed-choice quantum eraser was realized by Kim \textit{et al.}\ in 1999 \cite{PhysRevLett.84.1} in a double-slit interference experiment. A similar double-slit experiment involving entanglement of photon polarization was later performed by Walborn \textit{et al.}\ in 2002 \cite{PhysRevA.65.033818}.
More different scenarios framing the same concept have been experimentally realized  (see \cite{RevModPhys.88.015005} for a comprehensive review) including a recent work performed in a quantum circuit on the IBM Quantum platform \cite{chiou2022}.

Ever since the idea of quantum erasure was proposed, its interpretation and implication have been a subject of fierce controversy that continues to today \cite{englert1999quantum, mohrhoff1999objectivity, aharonov2005time, hiley2006erased, ellerman2015delayed, fankhauser2017taming, kastner2019delayed, qureshi2020} with divided opinions ranging from ``a magnificat affront to our conventional notions of space and time'' \cite{greene2004fabric} to ``an experiment that has caused no end of confusion'' \cite{carroll2019blog}.
Particularly, by analogy to the Einstein-Podolsky-Rosen-Bohm (EPR-Bohm) experiment \cite{PhysRev.108.1070,RevModPhys.81.1727}, Kastner argued that the quantum eraser neither erases nor delays any information, and does not present any mystery beyond the standard EPR correlation \cite{kastner2019delayed}.
Later on, by considering a Mach-Zehnder interferometer, which conveys the core idea of the quantum eraser more elegantly than a double-slit experiment, Qureshi further elaborated on the analogy between the quantum eraser and the EPR-Bohm experiment and claimed that there is no retrocausal effect whatsoever \cite{qureshi2020}.

In this paper, by generalizing the Mach-Zehnder interferometer considered in \cite{qureshi2020} with a nonsymmetric beam splitter (i.e., the transmission and reflection coefficients are not equal), we show that the quantum eraser shares \emph{exactly} the same formal (i.e., mathematical) structure with the EPR-Bohm experiment, as the modified Mach-Zehnder interferometer is \emph{exactly} analogous to a Stern-Gerlach apparatus used in the EPR-Bohm experiment. However, although the quantum eraser is \emph{formally} equivalent to the EPR-Bohm experiment, the former still raises \emph{conceptual} issues that cannot be explained out by the analogy to the latter, as opposed to what is claimed in \cite{kastner2019delayed,qureshi2020}. Specifically, if one applies \emph{counterfactual reasoning} about the which-way information, then whether a quanton travels along either of the two paths or both of them indeed can be affected by the delayed-choice measurement. This does not violate causality, because what can be altered is not the detection outcome but the which-way information, which is counterfactual in nature in most situations. The quantum eraser does present an additional conceptual ``mystery'' beyond the standard EPR puzzle, unless counterfactual reasoning is completely dismissed.

Furthermore, the quantum eraser experiments can be classified into two major categories: the \emph{entanglement quantum eraser} and the \emph{Scully-Dr{\"u}hl-type quantum eraser}.
The entanglement quantum eraser relies on the entanglement of some internal states between a pair of quantons (referred to as the \emph{signal} and \emph{idler} quantons), such as the experiment performed by Walborn \textit{et al.}\ \cite{PhysRevA.65.033818}, which involves the entanglement of polarization between a pair of photons.
In the entanglement quantum eraser, the which-way information of the signal quanton is ``recorded'' in terms of some \emph{internal state} of the idler quanton, which can be either read out or erased by different delayed-choice measurements.
However, as the which-way information of the signal quanton is inferred from the internal state of the idler quanton through counterfactual reasoning, one can always maintain that, without regard to counterfactual inference, the which-way information of the signal quanton is not recorded in the first place at all and not erased later.
On the other hand, in the Scully-Dr{\"u}hl-type quantum eraser as originally proposed by Scully and Dr{\"u}hl \cite{PhysRevA.25.2208} and performed by Kim \textit{et al.}\ \cite{PhysRevLett.84.1}, the which-way information of the signal quanton is ``recorded'' in terms of the states of two objects that are \emph{spatially separated}. The which-way information inferred from the measurement upon the states of the two objects becomes factual if the measurement yields a conclusive outcome that discriminates the record. Therefore, it makes good sense to say that the two objects serve as the ``recorders'' of the which-way information and the record of the which-way information is ``erased'' if the potentiality to yield a conclusive outcome is eliminated.
In this paper, we investigate the entanglement quantum eraser and the Scully-Dr{\"u}hl-type quantum eraser separately in depth. In terms of the Mach-Zehnder interferometer with a nonsymmetric beam splitter, we make it explicit that both kinds of quantum erasers are \emph{formally} equivalent to the EPR-Bohm experiment. Nevertheless, both raise \emph{conceptual} issues beyond the standard EPR puzzle, and the Scully-Dr{\"u}hl-type quantum eraser presents a ``mystery'' deeper than that of the entanglement quantum eraser.

We also consider the quantum eraser in view of the many-worlds interpretation (MWI) of quantum mechanics \cite{RevModPhys.29.454, dewitt1975many, RevModPhys.29.463}.
The MWI provides an appealing ontological framework, wherein all possible experimental outcomes exist simultaneously and thus many paradoxes of quantum mechanics (including the quantum eraser) are simply resolved as they are no longer matters of concern \cite{dewitt1970}.
Our analysis affirms that the standard (i.e., Copenhagen) interpretation and the MWI yield identical experimental predictions beyond a trivial model.
However, the investigation of the quantum eraser also reveals the subtle disharmony between the theoretical formulation of the MWI and its practical application: in principle the classical notion of being in a definite state is completely repudiated, but in practice it still has to resort to (semi)classical reasoning about definite states in order to theorize the dynamics of evolution.

This paper is organized as follows. In \secref{sec:MZ interferometer}, we consider the modified Mach-Zehnder interferometer with a nonsymmetric beam splitter, which draws a close analogy to the Stern-Gerlach apparatus. In \secref{sec:entanglement quantum eraser}, we study the entanglement quantum eraser in terms of the modified Mach-Zehnder interferometer and investigate its equivalence to and difference from the EPR-Bohm experiment. In \secref{sec:SD quantum eraser}, the same investigation is carried out for the Scully-Dr{\"u}hl-type quantum eraser. In \secref{sec:MWI}, we reconsider the quantum eraser in the MWI. Finally, the conclusions are summarized in \secref{sec:summary}.

\newpage

\section{Modified Mach-Zehnder interferometer}\label{sec:MZ interferometer}
\begin{figure}
\centering
    \includegraphics[width=0.7\textwidth]{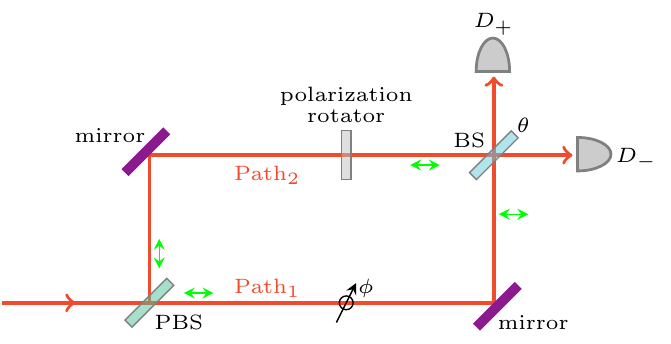}
\caption{The schematic diagram of a Mach-Zehnder interferometer that separates different polarizations into the two paths.}
\label{fig:interferometer}
\end{figure}

Consider the experimental setup using a Mach-Zehnder interferometer as sketched in \figref{fig:interferometer}, which couples the photon's spatial degree of freedom with its degree of polarization. An incident photon is split by the polarizing beam splitter $\mathrm{PBS}$ into two paths, $\Path{1}$ and $\Path{2}$, with horizontal ($\leftrightarrow$) and vertical ($\updownarrow$) polarizations, respectively.
Along $\Path{1}$, an adjustable phase shift $\phi$ is introduced (e.g, by inserting a phase-shift plate).
Along $\Path{2}$, a polarization rotator that rotates $\updownarrow$ into $\leftrightarrow$ is introduced in order to make the two paths interfere with each other.
The two paths are finally recombined by the beam splitter $\mathrm{BS}$ before the photon strikes either of the two detectors, $D_+$ and $D_-$.\footnote{Denoting the two detectors as $D_+$ and $D_-$, instead of $D_1$ and $D_2$, underscores the analogy to the Stern-Gerlach apparatus, as will be seen shortly.}
The detection probabilities at $D_+$ and $D_-$, which are measured as accumulated counts of repeated experiments of individual photons, are said to exhibit the two-path interference pattern if they appear as modulated in response to the phase shift $\phi$.

For generality, we consider $\mathrm{BS}$ to be a nonsymmetric beam splitter (i.e., the transmission and reflection coefficients are not equal). Supposing $\mathrm{BS}$ is lossless, we can describe the trasfer of $\mathrm{BS}$ by a $2\times2$ unitary matrix as
\begin{equation}
\left(
  \begin{array}{c}
    \ket{\Path{1}} \\
    \ket{\Path{2}} \\
  \end{array}
\right)
\mathop{\longrightarrow}\limits_\mathrm{BS}
\left(
  \begin{array}{cc}
    \alpha & \beta \\
    \beta^* & -\alpha^* \\
  \end{array}
\right)
\left(
  \begin{array}{c}
    \ket{D_+} \\
    \ket{D_-} \\
  \end{array}
\right),
\end{equation}
where $\abs{\alpha}^2$ and $\abs{\beta}^2$ are the transmission and reflection coefficients, respectively, which satisfy
\begin{equation}
\abs{\alpha}^2 + \abs{\beta}^2 = 1,
\end{equation}
$\ket{\Path{1/2}}$ represents the photon state localized on $\Path{1/2}$, and $\ket{D_\pm}$ represents the photon state that is going to strike $D_\pm$.
Because the arguments of the complex number $\alpha$ and $\beta$ can be absorbed into the phases of $\ket{D_+}$ and $\ket{D_-}$ as well as the phase shit $\phi$, we only need to consider the case that both $\alpha$ and $\beta$ are real. For convenience, we parameterize them by $\theta\in[0,\pi]$ as
\begin{equation}\label{alpha beta}
\alpha = \cos(\theta/2),
\qquad
\beta = \sin(\theta/2).
\end{equation}
The modified Mach-Zehnder interferometer captures the core concepts of the corresponding double-slit experiment.\footnote{In the double-slit experiment, we can place two polarizers with horizontal and vertical polarizations in front of the two slits to emulate the function of $\mathrm{PBS}$. A polarization rotator is then placed behind the second slit to make the paths from the two slits interfere with each other. The different positions on the screen correspond to different values of $\phi$. We can also place two optical attenuators with different attenuation rates behind the two slits to emulate different values of $\theta$.}

Given a horizontally polarized photon entering the interferometer, we denote its state as $\ket{\leftrightarrow}\ket{\psi}$, where $\ket{\psi}$ represents its spatial degree of freedom. The polarizing beam splitter $\mathrm{PBS}$ transfers $\ket{\psi}$ into $\ket{\Path{1}}$. The photon then undergoes the phase shift $\phi$ and $\mathrm{BS}$.
The state transfer of $\ket{\leftrightarrow}\ket{\psi}$ before hitting $D_+$ or $D_-$ is given by
\begin{subequations}\label{horizontal}
\begin{eqnarray}
\ket{\leftrightarrow}\ket{\psi}
& \mathop{\longrightarrow}\limits_\mathrm{PBS} & \ket{\leftrightarrow}\ket{\Path{1}} \\
& \mathop{\longrightarrow}\limits_{\phi} & e^{i\phi}\ket{\leftrightarrow}\ket{\Path{1}} \\
& \mathop{\longrightarrow}\limits_\mathrm{BS} &
\ket{\leftrightarrow} \left(e^{i\phi}\alpha\ket{D_+} + e^{i\phi}\beta\ket{D_-}\right).
\end{eqnarray}
\end{subequations}
Similarly, given a vertically polarized photon, denoted as $\ket{\updownarrow}\ket{\psi}$, it undergoes $\mathrm{PBS}$, the polarization rotator, and $\mathrm{BS}$ consecutively, and the state transfer is given by
\begin{subequations}\label{vertical}
\begin{eqnarray}
\ket{\updownarrow}\ket{\psi}
& \mathop{\longrightarrow}\limits_\mathrm{PBS} & \ket{\updownarrow}\ket{\Path{2}} \\
& \mathop{\longrightarrow}\limits_{\mathrm{rotator}} & \ket{\leftrightarrow}\ket{\Path{2}} \\
& \mathop{\longrightarrow}\limits_\mathrm{BS} &
\ket{\leftrightarrow} \left(\beta^*\ket{D_+} - \alpha^*\ket{D_-}\right).
\end{eqnarray}
\end{subequations}

Define two orthonormal basis states of arbitrary (elliptical) polarizations parameterized by $\vartheta$ and $\varphi$ in the ``spinor'' style as
\begin{subequations}\label{n + -}
\begin{eqnarray}
\ket{\hat{n}_{\vartheta,\varphi},+} &:=& \cos(\vartheta/2)\ket{\leftrightarrow} + e^{i\varphi} \sin(\vartheta/2)\ket{\updownarrow}
\equiv
\left(
  \begin{array}{c}
    \cos(\vartheta/2) \\
    e^{i\varphi} \sin(\vartheta/2) \\
  \end{array}
\right),\\
\ket{\hat{n}_{\vartheta,\varphi},-} &:=& \sin(\vartheta/2)\ket{\leftrightarrow} - e^{i\varphi} \cos(\vartheta/2)\ket{\updownarrow}
\equiv
\left(
  \begin{array}{c}
    \sin(\vartheta/2) \\
    -e^{i\varphi} \cos(\vartheta/2) \\
  \end{array}
\right).
\end{eqnarray}
\end{subequations}
By virtue of linearity upon \eqref{horizontal} and \eqref{vertical}, a photon with the polarization $\ket{\hat{n}_{\vartheta,\varphi},+}$ or $\ket{\hat{n}_{\vartheta,\varphi},-}$ entering the interferometer will undergo the transfer:
\begin{subequations}\label{evolution}
\begin{eqnarray}
\ket{\hat{n}_{\vartheta,\varphi},+}\ket{\psi}
& \longrightarrow &
\ket{\leftrightarrow}
\left(
\left(\alpha\cos(\vartheta/2)e^{i\phi} + \beta\sin(\vartheta/2)e^{i\varphi}\right)\ket{D_+}
\right. \nonumber\\
&& \qquad \quad \mbox{} +
\left.
\left(\beta^*\cos(\vartheta/2)e^{i\phi} - \alpha^*\sin(\vartheta/2)e^{i\varphi}\right)\ket{D_-}
\right), \nonumber\\
&\equiv&
e^{i\phi}\ket{\leftrightarrow}
\left(
\inner{\hat{n}_{\theta,\phi},+}{\hat{n}_{\vartheta,\varphi},+}\ket{D_+}
+
\inner{\hat{n}_{\theta,\phi},-}{\hat{n}_{\vartheta,\varphi},+}\ket{D_-}
\right), \\
\ket{\hat{n}_{\vartheta,\varphi},-}\ket{\psi}
& \longrightarrow &
\ket{\leftrightarrow}
\left(
\left(\alpha\sin(\vartheta/2)e^{i\phi} - \beta\cos(\vartheta/2)e^{i\varphi}\right)\ket{D_+}
\right. \nonumber\\
&& \qquad \qquad \mbox{} +
\left.
\left(\beta^*\sin(\vartheta/2)e^{i\phi} + \alpha^*\cos(\vartheta/2)e^{i\varphi}\right)\ket{D_-}
\right), \nonumber\\
&\equiv&
e^{i\phi}\ket{\leftrightarrow}
\left(
\inner{\hat{n}_{\theta,\phi},+}{\hat{n}_{\vartheta,\varphi},-}\ket{D_+}
+
\inner{\hat{n}_{\theta,\phi},-}{\hat{n}_{\vartheta,\varphi},-}\ket{D_-}
\right),
\end{eqnarray}
\end{subequations}
where we have used \eqref{alpha beta}.
Consequently, for a photon with arbitrary initial polarization $\ket{\hat{n}_{\vartheta,\varphi},\pm}$, the probability that it registers a signal in $D_+$ or $D_-$, respectively, is given by
\begin{subequations}\label{prob}
\begin{eqnarray}
P(D_+) &=& \abs{\inner{\hat{n}_{\theta,\phi},+}{\hat{n}_{\vartheta,\varphi},\pm}}^2, \\
P(D_-) &=& \abs{\inner{\hat{n}_{\theta,\phi},-}{\hat{n}_{\vartheta,\varphi},\pm}}^2.
\end{eqnarray}
\end{subequations}
The modified Mach-Zehnder interferometer as shown in \figref{fig:interferometer} bears a close resemblance to the Stern-Gerlach experiment for spin-$1/2$ particles.
The photon polarization states $\ket{\hat{n}_{\vartheta,\varphi},+}$ and $\ket{\hat{n}_{\vartheta,\varphi},-}$ defined in \eqref{n + -} are analogous to the spin-up and spin-down states (also denoted as $\ket{\hat{n}_{\vartheta,\varphi},\pm}$) along the orientation $\hat{n}_{\vartheta,\varphi}=(\sin\vartheta\cos\varphi, \sin\vartheta\sin\varphi, \cos\vartheta)$.
The interferometer parameterized by $\theta$ and $\phi$ is analogous to the Stern-Gerlach apparatus oriented in the direction $\hat{n}_{\theta,\phi}=(\sin\theta\cos\phi, \sin\theta\sin\phi, \cos\theta)$.
Signals registered in $D_+$ and $D_-$, respectively, correspond to the upper and lower traces on the screen in the Stern-Gerlach experiment.
Given an initial spin state $\ket{\hat{n}_{\vartheta,\varphi},\pm}$ entering the Stern-Gerlach apparatus oriented in the direction $\hat{n}_{\theta,\phi}$, the probability that it leaves an upper or lower trance on the screen is govern by the same equation as \eqref{prob}.\footnote{\label{foot:analogy}In the same spirit, the modified Mach-Zehnder interferometer parameterized by $\theta$ and $\phi$ is also analogous to a \emph{single} polarizing beam splitter that splits the incident beam into two beams of orthonormal polarizations $\ket{\hat{n}_{\theta,\phi},+}$ and $\ket{\hat{n}_{\theta,\phi},-}$. We focus on the Stern-Gerlach apparatus oriented in $\hat{n}_{\theta,\phi}$ as the representative example for the analogy, as it is easier to adjust both $\theta$ and $\phi$ for the Stern-Gerlach apparatus than for a polarizing beam splitter.}

In terms of the two unit vectors $\hat{n}_{\theta,\phi}$ and $\hat{n}_{\vartheta,\varphi}$, the inner product appearing in \eqref{prob} can be elegantly recast as
\begin{subequations}\label{inner product}
\begin{eqnarray}
\abs{\inner{\hat{n}_{\theta,\phi},+}{\hat{n}_{\vartheta,\varphi},\pm}}^2 &=& \frac{1\pm \hat{n}_{\theta,\phi}\cdot\hat{n}_{\vartheta,\varphi}}{2}, \\
\abs{\inner{\hat{n}_{\theta,\phi},-}{\hat{n}_{\vartheta,\varphi},\pm}}^2 &=& \frac{1\mp \hat{n}_{\theta,\phi}\cdot\hat{n}_{\vartheta,\varphi}}{2}.
\end{eqnarray}
\end{subequations}
Particularly, if the entering photon is in a polarization state with equal probability of being $\ket{\leftrightarrow}$ and $\ket{\updownarrow}$, i.e.,
\begin{equation}
\ket{\hat{n}_{\vartheta=\pi/2,\varphi},+}
=\frac{1}{\sqrt{2}}\left(\ket{\leftrightarrow} + e^{i\varphi}\ket{\updownarrow}\right),
\end{equation}
it follows from \eqref{inner product} that
\begin{equation}
P(D_\pm) = \frac{1\pm\sin\theta\cos(\phi-\varphi)}{2},
\end{equation}
where the modulation in response to the adjustable phase shift $\phi$ is a manifestation of the two-path interference.
The contrast of the interference pattern depends on $\theta$. It reaches maximum for the case of $\theta=\pi/2$ ( i.e., the beam splitter $\mathrm{BS}$ is symmetric).
At the other extreme, if $\theta=0$ (i.e., $\mathrm{BS}$ is completely transparent) or $\theta=\pi$ (i.e., $\mathrm{BS}$ is completely reflective), the interference pattern is completely diminished. This is because $\Path{1}$ and $\Path{2}$ strike $D+$ and $D_-$ separately, and thus the two paths do not interfere with each other at all.

If the entering photon is completely unpolarized, its polarization is described by the density matrix
\begin{equation}\label{unpolarized}
\rho=\frac{1}{2}\mathbbm{1}_{2\times2}.
\end{equation}
The detection probability is given by
\begin{equation}\label{flat prob}
P(D_\pm) = \Tr \left(\ket{\hat{n}_{\theta,\phi},\pm}\bra{\hat{n}_{\theta,\phi},\pm}\rho\right) = \frac{1}{2},
\end{equation}
which does not exhibit any interference pattern.

However, there is a subtle yet crucial issue that differentiates the modified Mach-Zehnder interferometer from the Stern-Gerlach apparatus (or the single polarizing beam splitter mentioned in \ftref{foot:analogy}).
The Stern-Gerlach apparatus oriented in $\hat{n}_{\theta,\phi}$ functions as a \emph{projective measurement} that projects an arbitrary spin state into the spin-up and spin-down states in the direction $\hat{n}_{\theta,\phi}$. In the same spirit, can we view the modified Mach-Zehnder interferometer parameterized by $\theta$ and $\phi$ \emph{as a whole} as a projective measurement that projects an entering photon state into $\ket{\hat{n}_{\theta,\phi},+}$ and $\ket{\hat{n}_{\theta,\phi},-}$?

Consider a photon polarized in $\ket{\leftrightarrow}$ entering a modified Mach-Zehnder interferometer parameterized by $\theta$ and $\phi$. Since
\begin{equation}\label{special case}
\ket{\leftrightarrow} = \cos(\theta/2)\ket{\hat{n}_{\theta,\phi},+} + \sin(\theta/2)\ket{\hat{n}_{\theta,\phi},-},
\end{equation}
this photon will click $D_+$ or $D_-$ with probability $\cos^2(\theta/2)$ or $\sin^2(\theta/2)$, respectively. If we assert that the initial state $\ket{\leftrightarrow}$ is collapsed into $\ket{\hat{n}_{\theta,\phi},+}:=\cos(\theta/2)\ket{\leftrightarrow} + e^{i\phi} \sin(\theta/2)\ket{\updownarrow}$ or $\ket{\hat{n}_{\theta,\phi},-}:=\sin(\theta/2)\ket{\leftrightarrow} - e^{i\phi} \cos(\theta/2)\ket{\updownarrow}$ according to the final signal in $D_\pm$,
then the photon is said to have traveled both paths (except for the case $\theta=0$ or $\theta=\pi$), which contradicts the proposition that the initial state $\ket{\leftrightarrow}$ travels only $\Path{1}$.
However, although as obvious as it seems, the proposition that $\ket{\leftrightarrow}$ travels only $\Path{1}$ and $\ket{\updownarrow}$ travels only $\Path{2}$ is counterfactual (i.e, unverifiable) in general and becomes factual only when $\theta=0$ or $\theta=\pi$ in particular.
That is, for the case that $\theta\neq0,\pi$, any claim about whether the photon travels along $\Path{1}$, $\Path{2}$, or both in a specific setting cannot be experimentally verified without modifying $\theta$ or $\phi$ or employing some ``path detectors'' along the paths.
Therefore, although as absurd as it may seem, if only verifiable factuality is concerned, it is counterintuitive but not fallacious at all to maintain the counterfactual conviction: as long as $\theta\neq0,\pi$, the initial states $\ket{\leftrightarrow}$ and $\ket{\updownarrow}$ travel both paths as well as other polarization states.
If one accepts this conviction, it is then perfectly legitimate to view the modified Mach-Zehnder interferometer parameterized by $\theta$ and $\phi$ as a whole as a projective measurement just like the Stern-Gerlach apparatus oriented in $\hat{n}_{\theta,\phi}$.
From this standpoint, the modified Mach-Zehnder interferometer is exactly analogous to the Stern-Gerlach apparatus, and consequently the mystery of the quantum eraser can be understood completely by the analogy to the EPR-Bohm experiment.
If, on the other hand, one adopts the different (but more intuitive) counterfactual conviction that $\ket{\leftrightarrow}$ travels $\Path{1}$ and $\ket{\updownarrow}$ travels $\Path{2}$ exclusively, then the fact that the photon of $\ket{\leftrightarrow}$ or $\ket{\updownarrow}$ clicks $D_\pm$ probabilistically with the probability given by \eqref{prob} is said to be caused by the beam splitter $\mathrm{BS}$ \emph{alone}, not by the interferometer \emph{as a whole}.
According to this conviction, the delayed-choice quantum eraser does raise a conceptual puzzle beyond the standard EPR paradox.
We will elaborate on the mystery of the quantum eraser in view of the EPR-Bohm experiment in the next section.

\section{Entanglement quantum eraser}\label{sec:entanglement quantum eraser}
Using the modified Mach-Zehnder interferometer in \figref{fig:interferometer}, we can conceive an entanglement quantum eraser that makes explicit the analogy to the EPR-Bohm experiment. This allows us to unravel the mystery of quantum erasure from the perspective of the well-understood EPR paradox.

\subsection{The setup}\label{sec:setup}
\begin{figure}
\centering
    \includegraphics[width=0.85\textwidth]{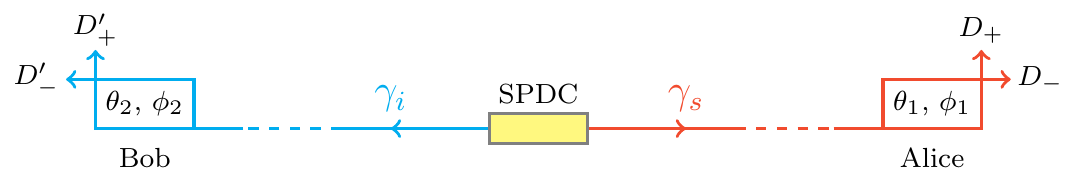}
\caption{A pair of entangled photons $\gamma_s$ and $\gamma_i$ with orthogonal polarization are created by SPDC. They are directed into the two modified Mach-Zehnder interferometers measured by Alice and Bob, respectively.}
\label{fig:two interferometers}
\end{figure}

Consider the experimental setup as sketched in \figref{fig:two interferometers}. Spontaneous parametric down-conversion (SPDC) in a nonlinear optical crystal is used to prepare a pair of entangled photons ($\gamma_s$ and $\gamma_i$) that are orthogonally polarized. The signal photon $\gamma_s$ is directed into the modified Mach-Zehnder interferometer parameterized by $\theta_1$ and $\phi_1$ with the detectors $D_+$ and $D_-$ measured by Alice. On the other hand, the idler photon $\gamma_i$ is directed into the other modified Mach-Zehnder interferometer parameterized by $\theta_2$ and $\phi_2$ with the detectors $D'_+$ and $D'_-$ measured by Bob.
Bob's interferometer can be located more distant away from the SPDC source than Alice's, so that by the time the idler photon $\gamma_i$ enters Bob's interferometer the signal photon $\gamma_s$ has already struck $D_+$ or $D_-$. Bob might even set the values of $\theta_2$ and $\phi_2$ in the ``delayed-choice'' manner --- i.e., $\theta_2$ and $\phi_2$ can be tuned \emph{after} $\gamma_i$ has registered a signal at $D_+$ or $D_-$.

The initial state of the entangled pair of photons is given by
\begin{equation}\label{EPR pair}
\ket{\Psi_0} = \frac{1}{\sqrt{2}}
\left(\ket{\leftrightarrow}_{\gamma_s}\ket{\updownarrow}_{\gamma_i} -\ket{\updownarrow}_{\gamma_s}\ket{\leftrightarrow}_{\gamma_i}\right)
\otimes\ket{\psi}_{\gamma_s}\ket{\psi}_{\gamma_i},
\end{equation}
where $\ket{\psi}_{\gamma_s}$ and $\ket{\psi}_{\gamma_i}$ represent the spatial degrees of freedom for $\gamma_s$ and $\gamma_i$, respectively.
Equivalently, $\ket{\Psi_0}$ can be recast in terms of arbitrary orthonormal polarization states as
\begin{equation}\label{EPR pair 2}
\ket{\Psi_0} = -\frac{e^{i\varphi}}{\sqrt{2}}
\left(\ket{\hat{n}_{\vartheta,\varphi},+}_{\gamma_s}\ket{\hat{n}_{\vartheta,\varphi},-}_{\gamma_i}
-\ket{\hat{n}_{\vartheta,\varphi},-}_{\gamma_s}\ket{\hat{n}_{\vartheta,\varphi},+}_{\gamma_i}\right)
\otimes\ket{\psi}_{\gamma_s}\ket{\psi}_{\gamma_i}.
\end{equation}
From Alice's viewpoint, the signal photon $\gamma_s$ is described by the reduced density matrix traced out over the degrees of the idler photon $\gamma_i$ as given by
\begin{equation}
\rho_0^{(\gamma_s)} = \Tr_{\gamma_i}\left(\ket{\Psi_0}\bra{\Psi_0}\right) = \frac{1}{2}\mathbbm{1}_{2\times2} \otimes\ket{\psi}_{\gamma_s}{}_{\gamma_s}\!\bra{\psi}.
\end{equation}
The polarization part is the same as \eqref{unpolarized} and therefore the probability that $\gamma_s$ clicks $D_+$ or $D_-$ is given by \eqref{flat prob}, showing no interference between the two paths (i.e., independent of $\phi_1$).
This may be understood as follows: as $\rho^{(\gamma_s)}$ can be interpreted as being either $\ket{\leftrightarrow}$ or $\ket{\updownarrow}$ by a fifty-fifty chance, for each individual event of the accumulated ensemble of repeated experiments, the photon $\gamma_s$ can be said to travel \emph{either} $\Path{1}$ \emph{or} $\Path{2}$ with equal probability.\footnote{\label{foot:unitary freedom}However, because of the \emph{unitary freedom} for density matrices, $\rho^{(s)}$ also admits infinitely many different interpretations. Therefore, it is only an interpretation, not an objective reality, to assert that $\gamma_s$ travels either $\Path{1}$ or $\Path{2}$ with equal probability. We will come back to this point shortly.}

Furthermore, if we also take into consideration the measurement by Bob, the accumulated events of $\gamma_s$ measured by $D_+$ and $D_-$ can be grouped into two subensembles according to whether $\gamma_i$ clicks $D'_+$ or $D'_-$.
According to \eqref{EPR pair 2}, from Bob's viewpoint, when $\gamma_i$ clicks $D'_+$ or $D'_-$, the polarization state of the pair of photons is collapsed into
\begin{equation}
\ket{\Psi_{D'_\pm}}=\ket{\hat{n}_{\theta_2,\phi_2},\mp}_{\gamma_s}\ket{\hat{n}_{\theta_2,\phi_2},\pm}_{\gamma_i}.
\end{equation}
Consequently, within the confines of the subensemble associated with $D'_+$ or $D'_-$, the detection probabilities of $D_+$ and $D_-$ are given by
\begin{subequations}\label{prob sub}
\begin{eqnarray}
P(D_\pm|D'_+) &=& \abs{\inner{\hat{n}_{\theta_1,\phi_1},\pm}{\hat{n}_{\theta_2,\phi_2},-}}^2 = \frac{1\mp \hat{n}_{\theta_1,\phi_1}\cdot\hat{n}_{\theta_2,\phi_2}}{2},\\
P(D_\pm|D'_-) &=& \abs{\inner{\hat{n}_{\theta_1,\phi_1},\pm}{\hat{n}_{\theta_2,\phi_2},+}}^2 = \frac{1\pm \hat{n}_{\theta_1,\phi_1}\cdot\hat{n}_{\theta_2,\phi_2}}{2}.
\end{eqnarray}
\end{subequations}
If Bob chooses $\theta_2=0$ (i.e., $\mathrm{BS}$ is completely transparent), \eqref{prob sub} becomes
\begin{subequations}
\begin{eqnarray}
P(D_\pm|D'_+) &=& \frac{1\mp \cos\theta_1}{2},\\
P(D_\pm|D'_-) &=& \frac{1\pm \cos\theta_1}{2}.
\end{eqnarray}
\end{subequations}
If Bob chooses $\theta_2=\pi$ (i.e., $\mathrm{BS}$ is completely reflective), \eqref{prob sub} becomes
\begin{subequations}
\begin{eqnarray}
P(D_\pm|D'_+) &=& \frac{1\pm \cos\theta_1}{2},\\
P(D_\pm|D'_-) &=& \frac{1\mp \cos\theta_1}{2}.
\end{eqnarray}
\end{subequations}
In the case of $\theta_2=0$ or $\theta_2=\pi$, whether the polarization of each individual $\gamma_s$ is $\ket{\leftrightarrow}$ or $\ket{\updownarrow}$ can be inferred from whether $\gamma_i$ clicks $D'_+$ or $D'_-$. If one adopts the intuitive conviction that $\ket{\leftrightarrow}$ travels $\Path{1}$ and $\ket{\updownarrow}$ travels $\Path{2}$ exclusively, then the ``which-way'' information of whether $\gamma_s$ travels along $\Path{1}$ or $\Path{2}$ is ``marked'' by the outcome of $D'_+$ or $D'_-$. As $\gamma_s$ is said to travel \emph{either} $\Path{1}$ \emph{or} $\Path{2}$, within each subensemble associated with $D'_+$ or $D'_-$, the detection probabilities of $D_+$ and $D_-$ exhibit no interference between the two paths.

On the other hand, if Bob chooses $\theta_2=\pi/2$ (i.e., $\mathrm{BS}$ is symmetric), \eqref{prob sub} becomes
\begin{subequations}\label{max erasure}
\begin{eqnarray}
P(D_\pm|D'_+) &=& \frac{1\mp \sin\theta_1\cos(\phi_1-\phi_2)}{2},\\
P(D_\pm|D'_-) &=& \frac{1\pm \sin\theta_1\cos(\phi_1-\phi_2)}{2},
\end{eqnarray}
\end{subequations}
which manifests the modulation in response to $\phi_1$ (provided $\theta_1\neq0,\pi$).
The which-way information of each individual $\gamma_s$ is completely ``unmarked'' by the outcome of $D'_1$ or $D'_2$, and $\gamma_s$ is said to travel \emph{both} $\Path{1}$ and $\Path{2}$ simultaneously. Therefore, within each subensemble associated with $D'_+$ or $D'_-$, the detection probabilities of $D_1$ and $D_2$ manifest the interference between the two paths.

Furthermore, If Bob chooses $\theta_2$ to some value different from $0$, $\pi$, or $\pi/2$, the which-way information of each individual $\gamma_s$ is partially marked to a certain degree. Accordingly, within each subensemble associated with $D'_+$ or $D'_-$, the detection probabilities of $D_+$ and $D_-$ appear partially modulated in response to $\phi_1$. That is, each subensemble does manifest the two-path interference, but the visibility of the interference pattern is diminished to a certain extent compared to that of the case of $\theta_2=\pi/2$.
A detailed analysis of the complementarity relations between interference visibility and which-way distinguishability for an entanglement quantum eraser can be found in \cite{chiou2022}.\footnote{We assume  maximum entanglement between $\gamma_s$ and $\gamma_i$ in this paper, whereas the analysis in \cite{chiou2022} also considers the extension that the degree of entanglement is adjustable.}

Whether $\gamma_s$ travels along $\Path{1}$, $\Path{2}$, or both seems to be influenced by the value of $\theta_2$ chosen by Bob. This gives a rather astonishing implication: the behavior of $\gamma_s$ in the past apparently can be \emph{retroactively} affected by the choice made by Bob in the future.
In the literature, it is often claimed that the which-way information of each individual $\gamma_s$, which is often innocuously presupposed to be either $\Path{1}$ or $\Path{2}$, can be retroactively ``erased'' (to a certain degree depending on $\theta_2$) by the ``delayed choice'' made by Bob. As a result, the interference pattern of $\gamma_s$ is ``recovered'' within either subensemble associated with $D'_+$ or $D'_-$.

\subsection{In comparison to the EPR-Bohm experiment}\label{sec:comparison to EPR}
\begin{figure}
\centering
    \includegraphics[width=0.8\textwidth]{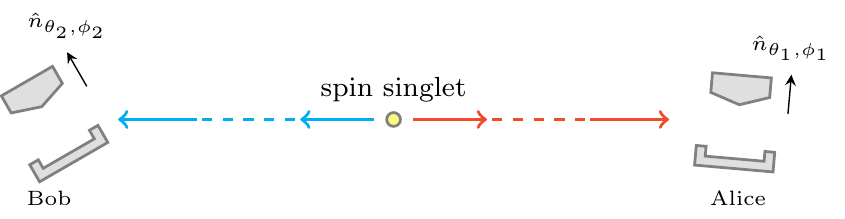}
\caption{The two entangled spin particles of a spin-singlet pair are directed into the two Stern-Gerlach apparati measured by Alice and Bob, respectively.}
\label{fig:EPR}
\end{figure}

As discussed in the last paragraph in \secref{sec:MZ interferometer}, if one adopts the conviction that, in the case of $\theta\neq0,\pi$, the initial states $\ket{\leftrightarrow}$ and $\ket{\updownarrow}$ travel both paths as well as other polarization states, then the modified Mach-Zehnder interferometer parameterized by $\theta$ and $\phi$ is exactly analogous to the Stern-Gerlach apparatus oriented in $\hat{n}_{\theta,\phi}=(\sin\theta\cos\phi, \sin\theta\sin\phi, \cos\theta)$.
Consequently, the experiment of the entanglement quantum eraser as depicted in \figref{fig:two interferometers} can be fully understood by the analogy to the EPR-Bohm experiment \cite{PhysRev.108.1070,RevModPhys.81.1727} as depicted in \figref{fig:EPR}.
Suppose a source emits pairs of spin-$1/2$ particles entangled in the spin singlet state. One particle of the pair is directed into the Stern-Gerlach apparatus oriented in $\hat{n}_{\theta_1,\phi_1}$ measured by Alice, whereas the other particle is directed into the Stern-Gerlach apparatus oriented in $\hat{n}_{\theta_2,\phi_2}$ measured by Bob.
Because of the spin-singlet entanglement, the outcomes (upper/lower traces, denoted as $D_\pm$) measured by Alice and the outcomes (upper/lower traces, denoted as $D'_\pm$) measured by Bob are correlated via the conditional probabilities
\begin{subequations}\label{prob EPR}
\begin{eqnarray}
P(D_\pm|D'_+) &=& P(D'_+|D_\pm) = \abs{\inner{\hat{n}_{\theta_1,\phi_1},\pm}{\hat{n}_{\theta_2,\phi_2},-}}^2 = \frac{1\mp \hat{n}_{\theta_1,\phi_1}\cdot\hat{n}_{\theta_2,\phi_2}}{2},\\
P(D_\pm|D'_-) &=& P(D'_-|D_\pm) = \abs{\inner{\hat{n}_{\theta_1,\phi_1},\pm}{\hat{n}_{\theta_2,\phi_2},+}}^2 = \frac{1\pm \hat{n}_{\theta_1,\phi_1}\cdot\hat{n}_{\theta_2,\phi_2}}{2},
\end{eqnarray}
\end{subequations}
which take the same form of \eqref{prob sub}.

In view of the analogy to the EPR-Bohm experiment, the fact that the interference pattern (i.e., modulation in response to $\phi_1$) is recovered within each subensemble associated with $D'_+$ or $D'_-$ is nothing but the \emph{EPR correlation} between $\ket{\hat{n}_{\theta_1,\phi_1},\pm}$ and $\ket{\hat{n}_{\theta_2,\phi_2},\pm}$ in disguise.
Furthermore, the EPR correlation is independent of whether Bob performs his measurement before or after Alice. The fact that Bob performs his measurement in the delayed-choice manner is not essential. In fact, Alice and Bob are completely on the equal footing --- on Bob's side, the interference pattern (i.e., modulation in response to $\phi_2$) can also be recovered within the subensembles associated $D_+$ and $D_-$ according to Alice's outcomes. That is, as shown in \eqref{prob EPR}, the conditional probability of $D_\pm$ given $D'_\pm$ is identical to that of $D'_\pm$ given $D_\pm$.\footnote{In the literature of quantum erasure, the extension with a variable $\phi_2$ on Bob's side is seldom considered. Thus, unfortunately, the fact that Alice and Bob are on equal footing is obscured and not often noted.}
In this sense, it has been argued that the quantum eraser does not erase any information at all, and the idea of erasure turns out to be inappropriate and misleading \cite{kastner2019delayed, qureshi2020}.
The claim that the which-way information of $\gamma_s$ is erased by the delayed choice made by Bob is based on the presupposition that $\gamma_s$ travels either $\Path{1}$ or $\Path{2}$ when it enters Alice's interferometer, but this ostensibly natural presupposition is only an interpretation among infinitely many others (recall \ftref{foot:unitary freedom}).

On the other hand, if one adopts the more intuitive conviction that $\ket{\leftrightarrow}$ travels $\Path{1}$ and $\ket{\updownarrow}$ travels $\Path{2}$ exclusively, then the which-way information of whether $\gamma_s$ travels $\Path{1}$, $\Path{2}$, or both is indeed influenced by the choice made by Bob as discussed earlier in \secref{sec:setup}.
As noted in the last paragraph in \secref{sec:MZ interferometer}, the which-way information of $\gamma_s$ is counterfactual in nature except for the case $\theta_1=0$ or $\theta_1=\pi$. Yet, ``counterfactual reasoning is not illegitimate per se.\,\dots [I]t is practiced daily, \dots'' (see Section 6-4 in \cite{peres1995quantum}).
If one employs the counterfactual reasoning in conformity to our intuition about which-way information, then whether $\gamma_s$ travels $\Path{1}$, $\Path{2}$, or both indeed can be affected by the choice made by Bob, even retroactively.
Therefore, unless one dismisses all counterfactual reasoning, the quantum eraser does present an additional ``mystery'' beyond the standard EPR puzzle, as opposed to what is claimed in \cite{kastner2019delayed,qureshi2020}.
It is agreeable that the notion of ``erasure'' is misleading as advocated in \cite{kastner2019delayed,qureshi2020}, but it would be too dismissive to relegate the quantum eraser to merely a manifestation of the EPR correlation and nothing else.
Furthermore, when it comes to the Scully-Dr{\"u}hl-type interferometer, the ``mystery'' beyond the standard EPR-Bohm experiment becomes even more prominent and the notion of ``erasure'' actually makes good sense, as will be seen in \secref{sec:SD quantum eraser}.

Also note that, because the quantum eraser as depicted in \figref{fig:two interferometers} is \emph{formally} equivalent to the EPR-Bohm experiment as depicted in \figref{fig:EPR}, we can formulate the Bell inequality or, more specifically, the CHSH inequality (see e.g., \cite{peres1995quantum}) for the quantum eraser in the obvious way as for the EPR-Bohm experiment, and show that the inequality can be violated. This implies that any local hidden-variable theories for the quantum eraser are incompatible with the experimental results.

\subsection{Quantum collapse and consistent histories}
In the standard EPR-Bohm experiment, there arises a perplexing puzzle: who, Alice or Bob, causes the entangled state to collapse in the first place? This puzzle has been addressed in \cite{Chiou_2013}.

The entangled state used in the EPR-Bohm experiment is given by the same form as \eqref{EPR pair 2} except that $\ket{\hat{n}_{\vartheta,\varphi},\pm}$ now represents spin state instead of polarization state.
From Alice's standpoint, the spin singlet is collapsed into $\ket{\hat{n}_{\theta_1,\phi_1},\pm}_s\ket{\hat{n}_{\theta_1,\phi_1},\mp}_i$ upon the moment when the particle flying toward Alice's apparatus strikes the screen and leaves an upper or lower trace.
The outcome of Bob's measurement then is predicted by Alice in terms of the probability given by \eqref{prob EPR}.
On the other hand, from Bob's standpoint, the spin singlet is collapsed into $\ket{\hat{n}_{\theta_2,\phi_2},\mp}_s\ket{\hat{n}_{\theta_2,\phi_2},\pm}_i$ upon the moment when the particle flying toward Bob's apparatus strikes the screen and leaves an upper or lower trace.
The outcome of Alice's measurement then is predicted by Bob in terms \eqref{prob EPR} as well.
So, who collapses the entangled state into a product state in the first place?

The EPR correlation is independent of whether Alice performs her measurement before or after Bob. Furthermore, in the context of special relativity, if Alice and Bob perform their measurements at two spacetime events that are spacelike separated, the time-ordering of the two events can flip under a Lorentz boost. Therefore, the time-ordering of the two measurement has no physical significance in regard to the EPR correlation.
Alice and Bob can both claim that the entangled state is collapsed into a product state by her/his measurement, yet the predictions by Alice and Bob are consistent to each other.
In the case that Alice performs her measurement before Bob and the two events are causally related (thus the time-ordering of the two events cannot be flipped), Bob can still maintain that the entangled state is collapsed by his measurement, and accordingly he can \emph{retrodict} Alice's measurement. Alice's prediction about Bob
and Bob's retrodiction about Alice again lead to no inconsistency.

At first thought, it appears nonsensical that quantum mechanics can be used for retrodiction as well as for prediction, since this would imply that the collapse caused by a measurement made in the present can affect states not only in the future but also in the past.
It turns out this constitutes no violation against casuality. In fact, this is exactly what happens in the Wheeler's delayed-choice experiment \cite{wheeler1983}, which has been confirmed in various experimental realizations \cite{alley1986delayed, PhysRevA.35.2532, baldzuhn1989wave, PhysRevA.54.5042, KAWAI1998259, jacques2007experimental}.
The aspect of consistency between prediction and retrodiction has been discussed in \cite{PhysRevD.66.023510} and is closely related to the ``consistent histories'' interpretation of quantum mechanics \cite{PhysRevLett.75.3038}.
In a word, Alice and Bob can claim their own interpretations of what has happened or what will happen. Their interpretations may be different from each other but do not result in any inconsistency as far as verifiable factuality is concerned.

The quantum eraser manifests the aspect of having different yet consistent histories even further than the EPR-Bohm experiment.
When the signal photon $\gamma_s$ clicks $D_+$ or $D_-$, Alice preferably (although not necessarily) deduces that $\gamma_s$ has been in the initial state $\ket{\hat{n}_{\theta_1,\phi_1},\pm}$. As long as $\theta_1\neq0,\pi$, she then preferably theorizes the which-way history of $\gamma_s$ to be having traveled \emph{both} paths.
Consider the situation that Bob tunes $\theta_2=0$ and performs his measurement in a delayed moment later. When the idler photon $\gamma_i$ clicks $D'_+$ or $D'_-$, Bob necessarily deduces that $\gamma_i$ has been in the initial state $\ket{\leftrightarrow}$ or $\ket{\updownarrow}$, and preferably infers that $\gamma_s$ has been in the initial state $\ket{\updownarrow}$ or $\ket{\leftrightarrow}$, correspondingly. He then preferably theorizes the which-way history of $\gamma_s$ to be having traveled \emph{only} $\Path{2}$ or \emph{only} $\Path{1}$, respectively.
Therefore, Alice and Bob have different interpretations about the history of $\gamma_s$, yet both interpretations are perfectly compatible with the outcomes of $D_\pm$ and $D'_\pm$ and the correlation between them.
As noted earlier, the which-way history of $\gamma_s$ is counterfactual in nature except for the case $\theta_1=0$ or $\theta_1=\pi$, and similarly the which-way history of $\gamma_i$ is counterfactual in nature except for the case $\theta_2=0$ or $\theta_2=\pi$. Therefore, as far as verifiable actuality is concerned, it is perfectly legitimate for Alice and Bob to theorize different counterfactual interpretations about whether $\gamma_s$ has traveled $\Path{1}$, $\Path{2}$, or both.\footnote{In \cite{qureshi2021delayed}, it is claimed that ``the delayed-choice quantum eraser leaves no choice'' in a particular situation that is equivalent to the configuration with $\theta_1=\pi/2$, $\phi_1=\pi/2$ considered in this paper. In this situation, because each individual $\gamma_s$ registered in $D_\pm$ corresponds to the initial state $\ket{\hat{n}_{\theta_1=\pi/2,\phi_1=\pi/2}} \equiv\frac{1}{\sqrt{2}}\left(\ket{\leftrightarrow}\pm i\ket{\updownarrow}\right)$, which is commonly said to have traveled both paths equally, it is argued in \cite{qureshi2021delayed} that Bob hence ``no longer has the choice to seek either which-path information or quantum eraser''. This argument is based on a common fallacy that the history of $\gamma_s$ is factual and cannot be compatible with alternative descriptions. In fact, Bob always have the choice to yield a different history of $\gamma_s$ unless Alice sets $\theta_1=0$ or $\theta_1=\pi$, only for which the history of $\gamma_s$ becomes factual.}

\section{Scully-Dr{\"u}hl-type quantum eraser}\label{sec:SD quantum eraser}
As mentioned earlier, the idea that the which-way information of $\gamma_s$ can be erased by the delayed-choice measurement may be misleading (see \cite{kastner2019delayed, qureshi2020} for more arguments). It is suggested in \cite{kastner2019delayed} that the reason the notion of ``erasure'' of ``which way'' or ``both ways'' is introduced ``may be due to the fact that these spatial properties directly affect our perceptions by creating visual patterns,'' with which ``[w]e then identify the concept of `information about an observable'.''
However, we have argued that, although the quantum eraser bears exactly the same mathematical structure as the EPR-Bohm experiment, the former still raises a conceptual issue not to be found in the latter.
Especially, in a delay-choice quantum eraser of the type as originally proposed by Scully by Dr{\"u}hl \cite{PhysRevA.25.2208} and experimentally realized by Kim \textit{et al.}\ \cite{PhysRevLett.84.1}, the notion of ``erasure'' of which-way information does not arise merely from the visual patterns but makes good sense in line with semiclassical reasoning. In this section, we investigate the conceptual issues of the Scully-Dr{\"u}hl-type quantum eraser in more depth.

\subsection{The setup}
\begin{figure}
\centering
    \includegraphics[width=0.85\textwidth]{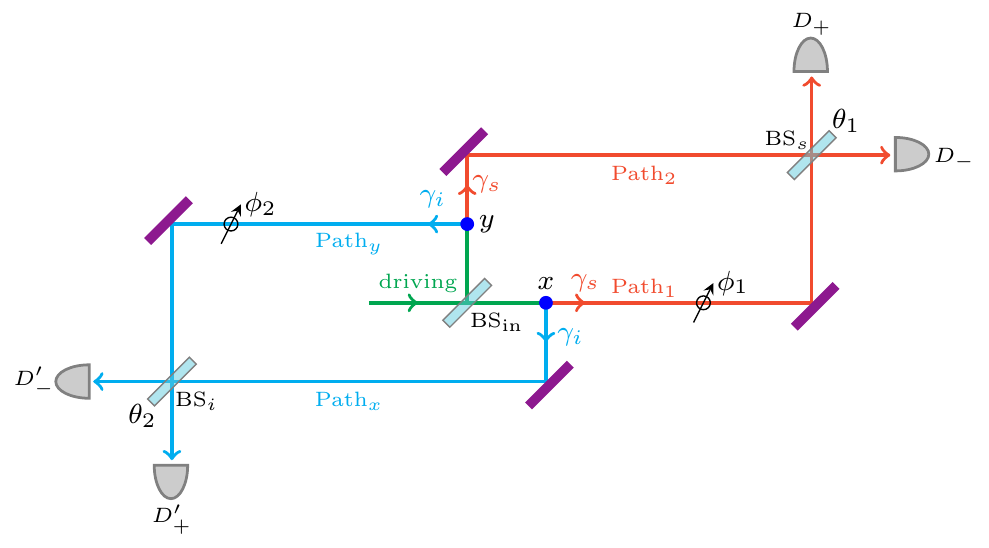}
\caption{The schematic diagram of the Scully-Dr{\"u}hl-type quantum eraser using two Mach-Zehnder interferometers, parameterized by $\theta_1$, $\phi_1$ and $\theta_2$, $\phi_2$, respectively.}
\label{fig:SD interferometer}
\end{figure}

\begin{figure}
\centering
    \includegraphics[width=0.95\textwidth]{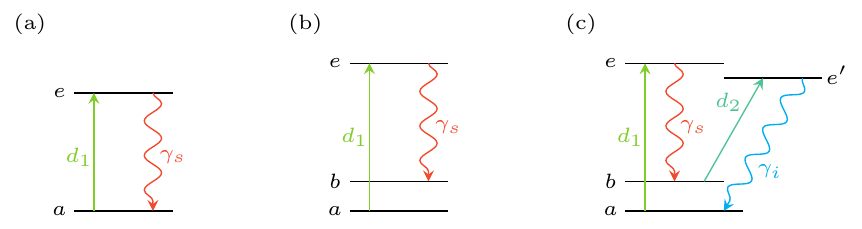}
\caption{Three different scenarios for the atoms placed at $x$ and $y$.
\textbf{(a)} The incident pulse $d_1$ excites $\ket{a}$ to $\ket{e}$. The state $\ket{e}$ emits a photon $\gamma_s$ and returns back to $\ket{a}$.
\textbf{(b)} The pulse $d_1$ excites $\ket{a}$ to $\ket{e}$. The state $\ket{e}$ emits a photon $\gamma_s$ and ends up at $\ket{b}$.
\textbf{(c)} A second pulse $d_2$ excites $\ket{b}$ to $\ket{e'}$. The state $\ket{e'}$ emits a second photon $\gamma_i$ and returns to $\ket{a}$.}
\label{fig:transition}
\end{figure}

To convey the core idea of the Scully-Dr{\"u}hl-type quantum eraser and to closely compare it with the EPR-Bohm experiment, we reformulate the double-slit experiment considered in the original proposal by Scully and Dr{\"u}hl \cite{PhysRevA.25.2208} into the setup using two Mach-Zehnder interferometers as sketched in \figref{fig:SD interferometer}.
Two atoms are located at $x$ and $y$ on $\Path{1}$ and $\Path{2}$, respectively. A driving photon pulse $d_1$ enters the symmetric beam splitter $\mathrm{BS_{in}}$ and impinges either of the two atoms.\footnote{In the Scully-Dr{\"u}hl-type quantum eraser, we do not consider the polarization degree of freedom for photons, and all beam splitters are non-polarizing ones, unlike $\mathrm{PBS}$ used in \figref{fig:interferometer}.}
The photon $d_1$ excites one of the atoms and trigger it to emit a photon $\gamma_s$, which travels along $\Path{1}$ and/or $\Path{2}$, then passes the beam splitter $\mathrm{BS}_s$ (parameterized by $\theta_1$), and finally strikes the detector $D_+$ or $D_-$ (measured by Alice). Along $\Path{1}$, an adjustable phase shift $\phi_1$ is introduced to form a modified Mach-Zehnder interferometer parameterized by $\theta_1$ and $\phi_1$. We consider three different scenarios of how the atom undergoes energy transition and emits $\gamma_s$ as depicted in \figref{fig:transition}.

In \figref{fig:transition} (a), we consider two-level atoms. The incident pulse $d_1$ excites one of the two atoms from its initial state $\ket{a}$ to the excited state $\ket{e}$. The excited atom then emits a photon $\gamma_s$ while returns back to the state $\ket{a}$. Because both atoms finally stay in $\ket{a}$, it is impossible to distinguish which atom emits $\gamma_s$. Correspondingly, the detection probabilities at $D_+$ and $D_-$ exhibits the two-path interference as modulated in response to the phase shift $\phi_1$.

In \figref{fig:transition} (b), we consider three-level atoms. The incident pulse $d_1$ excites one of the two atoms from its initial state $\ket{a}$ to the excited state $\ket{e}$. The excited atom then emits a photon $\gamma_s$ while transits to a different state $\ket{b}$. Here, the atom that emits $\gamma_s$ is finally in the state $\ket{b}$, while the other atom remains in the state $\ket{a}$. In a sense, the two atoms play the role as ``recorders'' that record the which-way information of $\gamma_s$. Since the which-way information has been ``recorded'' (and can be read out by measuring the internal states of the atoms), $\Path{1}$ and $\Path{2}$ do not interfere with each other. The detection probabilities at $D_+$ and $D_-$ does not exhibit any interference pattern.

In \figref{fig:transition} (c), we consider the same configuration as in (b). However, after $\gamma_s$ is emitted by one of the atoms, we shoot another photon pulse $d_2$ into $\mathrm{BS_{in}}$. The pulse $d_2$ excites the atom in $\ket{b}$ to another excited state $\ket{e'}$. The atom in $\ket{e'}$ then emits a second photon $\gamma_i$ while returns back to the state $\ket{a}$. The which-way information recorded in one of the atoms is transferred to $\gamma_i$. We then direct the photon $\gamma_i$ into the other Mach-Zehnder interferometer parameterized by $\theta_2$ and $\phi_2$ with the detectors $D'_+$ and $D'_-$ (measured by Bob) as depicted in \figref{fig:SD interferometer}. If Bob sets $\theta_2=0$ or $\theta_2=\pi$ (i.e., $\mathrm{BS}_i$ is completely transparent or reflective), the which-way information of $\gamma_s$ can be inferred from the outcome of $D'_+$ or $D'_-$. If, on the other hand, Bob sets $\theta_2=\pi/2$, then the which-way information is completely ``erased'', and correspondingly $\gamma_s$ is said to travel \emph{both} $\Path{1}$ and $\Path{2}$. Consequently, within each subensemble associated with $D'_+$ or $D'_-$, the detection probabilities of $D_1$ and $D_2$ manifest the interference between the two paths. Furthermore, If Bob sets $\theta_2$ to some value different from $0$, $\pi$, or $\pi/2$, the which-way information of each individual $\gamma_s$ is partially erased. Accordingly, within each subensemble associated with $D'_+$ or $D'_-$, the detection probabilities of $D_+$ and $D_-$ appear partially modulated in response to $\phi_1$, partially manifesting the two-path interference. The lengths of $\Path{x}$ and $\Path{y}$ can be made much longer than those of $\Path{1}$ and $\Path{2}$ so that Bob can choose the values of $\theta_2$ and $\phi_2$ after the photon $\gamma_s$ has struck $D_+$ or $D_-$. But it turns out whether Alice perform her measurement before or after Bob makes no difference in regard to the outcomes of $D_\pm$ and $D'_\pm$ and the correlation between them.

\subsection{Distinguishability and visibility}\label{sec:D and V}
In the experimental realization of the Scully-Dr{\"u}hl-type quantum eraser by Kim \textit{et al.}\ \cite{PhysRevLett.84.1}, in replace of the two atoms under the scenario in \figref{fig:SD interferometer} (c), SPDC is used to prepare a signal-idler pair of photons ($\gamma_s$ and $\gamma_i$) by shooting a weak laser beam into a nonlinear optical crystal BBO (beta barium borate).
In order to describe all possible situations, we schematically denote the process of producing the photon $\gamma_s$ from one of the two objects (atoms, nonlinear crystals, etc.) by injecting a driving pulse ($d_1$) or two pulses ($d_1$ and $d_2$) as
\begin{subequations}
\begin{eqnarray}
\ket{A}_x & \mathop{\longrightarrow}\limits_\text{driving} & \ket{B}_x\otimes\ket{\Path{1}}_{\gamma_s}, \\
\ket{A}_y & \mathop{\longrightarrow}\limits_\text{driving} & \ket{B}_y\otimes\ket{\Path{2}}_{\gamma_s},
\end{eqnarray}
\end{subequations}
where $\ket{\Path{1/2}}_{\gamma_s}$ represents the photon $\gamma_s$ on $\Path{1}$ or $\Path{2}$, $\ket{A}_{x/y}$ represents the initial state of the object at $x$ or $y$ and any additional coupled degrees of freedom, and $\ket{B}_{x/y}$ represents the final state of the object at $x$ or $y$ and any additional coupled degrees of freedom (which may include a second photon $\gamma_i$).
The process that the driving pulse or pulses enter the beam splitter $\mathrm{BS_{in}}$ and excite one of the two objects is given by
\begin{subequations}
\begin{eqnarray}
\text{driving pulse(s)} & \mathop{\longrightarrow}\limits_\mathrm{BS_{in}} &
\frac{1}{\sqrt{2}}\left(\ket{\Path{1}}_\mathrm{driving} + \ket{\Path{2}}_\mathrm{driving}\right) \\
\label{Psi0}
& \mathop{\longrightarrow}\limits_{x,\ y} &
\ket{\Psi_0}\equiv
\frac{1}{\sqrt{2}}
\left(
\ket{B}_x\ket{A}_y\ket{\Path{1}}_{\gamma_s} + \ket{A}_x\ket{B}_y\ket{\Path{2}}_{\gamma_s}
\right),
\end{eqnarray}
\end{subequations}
where $\ket{\Path{1/2}}_\mathrm{driving}$ represents the pulse(s) split into $\Path{1/2}$.
The photon state $\ket{\Path{1}}_{\gamma_s}$ and $\ket{\Path{2}}_{\gamma_s}$ then undergoes the transfer:
\begin{subequations}\label{evolution path}
\begin{eqnarray}
\ket{\Path{1}}_{\gamma_s} & \longrightarrow & e^{i\phi_1}\alpha\ket{D_+}_{\gamma_s} + e^{i\phi_1}\beta\ket{D_-}_{\gamma_s}, \\
\ket{\Path{2}}_{\gamma_s} & \longrightarrow & \beta^*\ket{D_+}_{\gamma_s} - \alpha^*\ket{D_-}_{\gamma_s},
\end{eqnarray}
\end{subequations}
where $\ket{D_{+/-}}_{\gamma_s}$ represents the photon state that will strike $D_{+/-}$, and $\alpha, \beta$ are given by \eqref{alpha beta} with $\theta$ replaced by $\theta_1$.
By virtue of linearity, before $\gamma_s$ hits the detectors, the driving pulse or pulses entering $\mathrm{BS_{in}}$ undergo the transfer:
\begin{eqnarray}\label{Psi}
\ket{\text{driving pulse(s)}}  \longrightarrow   \ket{\Psi}
&=&\frac{1}{\sqrt{2}}
\left(\ket{B}_x\ket{A}_y \left(e^{i\phi}\alpha\ket{D_+}_{\gamma_s} + e^{i\phi}\beta\ket{D_-}_{\gamma_s}\right)
\right. \nonumber\\
&&
\qquad \quad \mbox{}+
\left.\ket{A}_x\ket{B}_y \left(\beta^*\ket{D_+}_{\gamma_s} - \alpha^*\ket{D_-}_{\gamma_s}\right)
\right).
\end{eqnarray}

The reduced density matrix of $\gamma_s$ for the initial state $\ket{\Psi_0}$ in \eqref{Psi0} is obtained by tracing out the degrees of freedom other than $\gamma_s$, i.e.,
\begin{eqnarray}
\rho_0^{(\gamma_s)} &=& \Tr_{x\otimes y} \ket{\Psi_0}\bra{\Psi_0} \nonumber\\
&=& \frac{1}{2}\left(\ket{\Path{1}}\bra{\Path{1}}
+ \inner{A_x,B_y}{B_x,A_y} \ket{\Path{1}}\bra{\Path{2}} \right. \nonumber\\
&& \qquad \left.\mbox{} + \inner{B_x,A_y}{A_x,B_y} \ket{\Path{2}}\bra{\Path{1}} + \ket{\Path{2}}\bra{\Path{2}} \right),
\end{eqnarray}
where $\ket{A_x,B_y}\equiv\ket{A}_x\ket{B}_y$ and $\ket{B_x,A_y}\equiv\ket{B}_x\ket{A}_y$.
Following the definition used in \cite{PhysRevResearch.2.012031}, we define the \emph{purity} of $\rho_0^{(\gamma_s)}$ as
\begin{equation}\label{mu s}
\mu_s := \sqrt{2\Tr (\rho_0^{(\gamma_s)})^2-1}
=\abs{\inner{A_x,B_y}{B_xA_y}},
\end{equation}
which quantifies how pure the state of the source photon $\gamma_s$ is.\footnote{In the literature of quantum mechanics and especially quantum information, the purity of a matrix density $\rho$ is usually defined as $\gamma:=\Tr (\rho^2)$, which satisfies $1/d\leq\gamma\leq1$, where $d$ is the dimension of the Hilbert space. Here, we adopt a different definition used in \cite{PhysRevResearch.2.012031}, which is for the two-dimensional case and satisfies $0\leq\mu_s\leq1$.}

After the photon $\gamma_s$ is emitted, we can obtain the which-way information of $\gamma_s$ by immediately measuring whether $\gamma_s$ is in the state $\ket{\Path{1}}$ or $\ket{\Path{2}}$. Alternatively, instead of measuring $\gamma_s$ directly, we can measure whether the objects at $x$ and $y$ are in the state $\ket{\psi_1}\equiv\ket{B_x,A_y}$ or $\ket{\psi_2}\equiv\ket{A_x,B_y}$ and then deduce the which-way information correspondingly. This measurement might be done in a delayed-choice manner.
If $\ket{\psi_1}$ and $\ket{\psi_2}$ are not orthogonal to each other, it is impossible to discriminate them with complete certainty. The next best thing is to apply the \emph{unambiguous quantum state discrimination} (UQSD) \cite{anthony2000quantum, bergou2004quantum} on the two-dimensional Hilbert space $\mathcal{H}$ containing $\ket{\psi_1}$ and $\ket{\psi_2}$ with the positive operator-valued measure (POVM) given by the operators:
\begin{subequations}\label{Fs}
\begin{eqnarray}
F_1 &=& \frac{1}{1+\abs{\inner{\psi_1}{\psi_2}}}|\psi_2^\bot\rangle \langle\psi_2^\bot|, \\
F_2 &=& \frac{1}{1+\abs{\inner{\psi_1}{\psi_2}}}|\psi_1^\bot\rangle \langle\psi_1^\bot|, \\
F_? &=& I - F_x - F_y,
\end{eqnarray}
\end{subequations}
where $|\psi_1^\bot\rangle\in\mathcal{H}$ is the state orthogonal to $\ket{\psi_1}$ and $|\psi_2^\bot\rangle\in\mathcal{H}$ is the state orthogonal to $\ket{\psi_2}$.
If the outcome ``1'' is obtained, it is completely certain that the state is in $\ket{\psi_1}$ and correspondingly $\gamma_s$ is in $\ket{\Path{1}}$. If the outcome ``2'' is obtained, it is completely certain that the state is in $\ket{\psi_2}$ and $\gamma_s$ is in $\ket{\Path{2}}$. If the outcome ``?'' is obtained, however, it is inconclusive which one the state is. The UQSD method gives the highest probability of having a conclusive outcome as $P_\mathrm{conc}=1-\abs{\inner{\psi_1}{\psi_2}}$. The \emph{distinguishability} of which-way information is defined as\footnote{\label{foot:different D}The distinguishability $\mathcal{D}$ is defined in the same spirit as in \cite{coles2014equivalence, PhysRevA.93.062111}. It is different from the distinguishability defined in Equation (7) in \cite{PhysRevResearch.2.012031}, which bears a different meaning.}
\begin{equation}\label{D def}
\mathcal{D} := 2P_\mathrm{succ}-1,
\end{equation}
where $P_\mathrm{succ}$ is the probability of successfully guessing whether $\gamma_s$ is in $\ket{\Path{1}}$ or $\ket{\Path{2}}$. Obviously, we have $P_\mathrm{succ}=1/2*(1-P_\mathrm{conc})+1*P_\mathrm{conc}$, and consequently
\begin{equation}\label{D}
\mathcal{D} = P_\mathrm{conc}= 1-\mu_s.
\end{equation}
If $\ket{A}_x$ is identical to $\ket{B}_x$ and $\ket{A}_y$ is identical to $\ket{B}_y$, it is impossible to infer the which-way information by discriminating these two states. This corresponds to $\mu_s^2=1$ and $\mathcal{D}=0$. On the other extreme, if $\ket{A}_x$ is very different from $\ket{B}_x$ (more precisely, $\abs{\inner{A}{B}_x}\ll1$) or $\ket{A}_y$ is very different from $\ket{B}_y$ (more precisely, $|\inner{A}{B}_y|\ll1$), we can infer the which-way information with high certainty. This corresponds to $\mu_s^2\ll1$ and $\mathcal{D}\approx 1$. The which-way information is said to be recorded in terms of the state $\ket{B}_x$ or $\ket{B}_y$.

On the other hand, the reduced density matrix of $\gamma_s$ for the state $\ket{\Psi}$ in \eqref{Psi} is given by
\begin{eqnarray}
&&\quad\rho^{(\gamma_s)} = \Tr_{x\otimes y} \ket{\Psi}\bra{\Psi} =\nonumber\\
&& \frac{1}{2}\left(\ket{D_+}\bra{D_+}+\ket{D_-}\bra{D_-}\right) \\
&& \mbox{} + \frac{e^{i\phi_1}}{2}\inner{A_x,B_y}{B_x,A_y}
\left[\alpha\beta\ket{D_+}\bra{D_+}-\alpha^2\ket{D_+}\bra{D_-}
+\beta^2\ket{D_-}\ket{D_+}-\alpha\beta\ket{D_-}\ket{D_-}\right] \nonumber\\
&& \mbox{} + \frac{e^{-i\phi_1}}{2}\inner{B_x,A_y}{A_x,B_y}
\left[\alpha^*\beta^*\ket{D_+}\bra{D_+}+{\beta^*}^2\ket{D_+}\bra{D_-}
-{\alpha^*}^2\ket{D_-}\ket{D_+}-\alpha^*\beta^*\ket{D_-}\ket{D_-}\right].\; \nonumber
\end{eqnarray}
Expressing the complex number in the polar form:
\begin{equation}\label{polar form}
\inner{A_x,B_y}{B_xA_y} \equiv \mu_s e^{i\delta},
\end{equation}
we obtain the probabilities that the photon $\gamma_s$ clicks $D_+$ and $D_-$ as
\begin{equation}\label{SD interference}
P(D_\pm) = \Tr \left(\ket{D_\pm}\bra{D_\pm}\rho^{(\gamma_s)}\right) = \frac{1\pm\mu_s\sin\theta_1\cos(\phi_1+\delta)}{2}.
\end{equation}
The \emph{visibility} of the interference pattern is defined as and given by
\begin{equation}\label{V}
\mathcal{V} := \frac{\max_{\phi_1} P(D_\pm)-\min_{\phi_1} P(D_\pm)}
{\max_{\phi_1} P(D_\pm)+\min_{\phi_1} P(D_\pm)}
=\mu_s\abs{\sin\theta_1}.
\end{equation}

By \eqref{D} and \eqref{V}, we have the inequality relation
\begin{equation}\label{V D duality}
\mathcal{D}^2+\mathcal{V}^2 = (1-\mu_s)^2+ \mu_s^2\sin^2\theta_1 \leq (1-\mu_s^2) + \mu_s^2 \leq 1,
\end{equation}
which affirms the complementarity relation between interference visibility and which-way distinguishability \cite{PhysRevA.51.54, PhysRevLett.77.2154}.
In summary, the more distinguishable the which-way information is, the less visible the interference pattern appears. See \cite{coles2014equivalence, PhysRevA.93.062111} and references therein for more discussions on the complementarity between interference visibility and which-way distinguishability.

\subsection{Various examples}
In this subsection, we investigate various concrete examples for the complementarity relation between visibility and distinguishability.

Fist, consider the case that the final state of the object at $x$ or $y$ after emitting $\gamma_s$ is completely indistinguishable from its initial state. The scenario in \figref{fig:transition} (a) is a typical example. Formally, this case is described as
\begin{subequations}
\begin{eqnarray}
\ket{A}_x &=& \ket{B}_x = \ket{a_1}_x, \\
\ket{A}_y &=& \ket{B}_y = \ket{a_2}_y.
\end{eqnarray}
\end{subequations}
It follows from \eqref{polar form} that $\mu_s=1$, $\delta=0$, and consequently \eqref{SD interference} becomes
\begin{equation}\label{ordinary interference}
P(D_\pm) = \frac{1\pm\sin\theta_1\cos\phi_1}{2}.
\end{equation}
Because the which-way information is completely unmarked, \eqref{ordinary interference} yields the ordinary two-path interference pattern.

This scenario has been experimentally realized (in a double-slit experiment), first by Eichmann \textit{et al.}\ \cite{PhysRevLett.70.2359, PhysRevA.57.4176} using a laser beam to excite two trapped ${}^{198}\mathrm{Hg}^+$ ions, and more recently by Araneda \textit{et al.}\ \cite{PhysRevLett.120.193603} using two trapped ${}^{138}\mathrm{Ba}^+$ ions.
However, the observed interference patterns showed significantly lower visibility than the theoretical predication corresponding to \eqref{ordinary interference}. This is understood as a consequence of inevitable interaction with the remaining degrees of freedom of the atoms as well as with external systems and fields \cite{PhysRevResearch.2.012031}.

To take into account the remaining degrees of freedom of the atoms and the presence of external parties, the Hilbert space under consideration needs to be extended to include the additional degrees of freedom. This is done by replacing the states $\ket{A_x,B_y}$ and $\ket{B_x,A_y}$ with the extended ones as
\begin{subequations}\label{extension}
\begin{eqnarray}
\ket{A_x,B_y} &\dashrightarrow& \ket{a_1}_x\otimes\ket{a_2}_y\otimes\ket{m}, \\
\ket{B_x,A_y} &\dashrightarrow& \ket{a_1}_x\otimes\ket{a_2}_y\otimes\ket{n},
\end{eqnarray}
\end{subequations}
where
\begin{subequations}\label{m n}
\begin{eqnarray}
\ket{m}&=&\sum c_{m_x,m_y,m_E}\ket{m_x}\ket{m_y}\ket{m_E}, \\
\ket{n}&=&\sum d_{m_x,m_y,m_E}\ket{n_x}\ket{n_y}\ket{n_E},
\end{eqnarray}
\end{subequations}
represent the two after-emission states for the remaining degrees of the atoms at $x$ and $y$ and the degrees of external parties (indicated by the subscripts $x$, $y$, and $E$, respectively)  \cite{PhysRevResearch.2.012031}.
The results calculated in \secref{sec:D and V} still hold except that \eqref{extension} has to be substituted for $\ket{A_x,B_y}$ and $\ket{B_x,A_y}$.
With this substitution, \eqref{polar form} gives
\begin{equation}\label{m n product}
\mu_s e^{i\delta} = \inner{m}{n}.
\end{equation}
In reality, because the process of producing $\gamma_s$ inevitably leaves a footprint upon external systems as well as upon the remaining degrees of freedom of the atoms (e.g., recoil of the atoms due to absorption or emission of a photon), we have $\ket{m}\neq\ket{n}$ and therefore $\mu_s<1$. The visibility of the interference pattern given by \eqref{SD interference} is diminished by a factor of $\mu_s$. This explains why observed interference patterns show
significantly lower visibility.\footnote{\label{foot:real experiment}In most cases of real experiments, the remaining degrees and the degrees of external parties altogether should not be described by a pure state, but by a density matrix (most likely a thermal density matrix), because repeated runs of an experiment, although considered to be in the same setting, are in fact carried out in different environmental states due to thermal or other uncontrollable fluctuations. Therefore, the observed interference patten is the probabilistic average of \eqref{SD interference} averaged over the density matrix. Averaging the phase $\delta$ appearing in \eqref{SD interference} over the probabilities given by the density matrix will smear the modulation in response to $\phi_1$. As a result, the interference visibility is further diminished much more than merely by the factor of $\mu_s$.}

Next, consider the case that the final state of the object at $x$ or $y$ after emitting $\gamma_s$ is completely different from its initial state. The scenario in \figref{fig:transition} (b) is a typical example. Formally, this case is described as
\begin{subequations}\label{a b}
\begin{eqnarray}
\ket{A}_x &=& \ket{a_1}_x, \quad  \ket{B}_x = \ket{b_1}_x, \\
\ket{A}_y &=& \ket{a_2}_y, \quad  \ket{B}_y = \ket{b_2}_y.
\end{eqnarray}
\end{subequations}
More realistically, we can also take into account the remaining degrees of freedom and the states of external parties by employing the substitution:
\begin{subequations}\label{extension 2}
\begin{eqnarray}
\ket{A_x,B_y} &\dashrightarrow& \ket{a_1}_x\otimes\ket{b_2}_y\otimes\ket{m}, \\
\ket{B_x,A_y} &\dashrightarrow& \ket{b_1}_x\otimes\ket{a_2}_y\otimes\ket{n}.
\end{eqnarray}
\end{subequations}
As long $\inner{a_1}{b_1}_x=0$ or $\inner{a_2}{b_2}_y=0$ (i.e., $\ket{a_1}_x$ is completely different from $\ket{b_1}_x$ or $\ket{a_2}_y$ is completely different from $\ket{b_2}_y$), we have $\mu_s=0$ by \eqref{polar form}, and consequently \eqref{SD interference} becomes $P(D_+)=P(D_-)=1/2$. Because the which-way information is recorded in terms of $\ket{b_1}_x$ or $\ket{b_2}_y$, the interference pattern is completely diminished, whether additional degrees are taken into account or not.

The next case is the scenario in \figref{fig:transition} (c), but we investigate it in more depth in the next two subsections.

Additionally, we consider the experiment of two-path single-photon interferometry using the controllable entangled nonlinear bi-photon source (ENBS) as proposed and performed by Lee \textit{et al.}\ \cite{kyung2018frequency} and analyzed in depth by Yoon and Cho \cite{sciadv.abi9268}.
Instead of two atoms, two SPDC crystals at $x$ and $y$ are simultaneously seeded and pumped by two lasers to produce signal and idler photons. The two crystals are seeded by an ultra-narrow linewidth continuous-wave laser, thus providing two initial coherent states $\ket{\alpha_1}_x$ and $\ket{\alpha_2}_y$ of single-frequency photons at $x$ and $y$. A pump pulse from the other laser (a comb laser used to give frequency comb pulse trains in the real experiment \cite{kyung2018frequency}) is then injected into the interferometer to induce stimulated parametric-down conversion (StPDC) upon the two crystals. The StPDC process produces a signal photon $\gamma_s$ and at the same time elevates one of the two coherent states $\ket{\alpha_1}_x$ and $\ket{\alpha_2}_x$ to the single-photon-added coherent state (SPACS) $\ket{\alpha_1,1}_x$ or $\ket{\alpha_2,1}_y$. More precisely, the SPACSs are given by
\begin{equation}
\ket{\alpha_i,1}_{x/y} := \frac{1}{\sqrt{1+\abs{\alpha_i}^2}}\, \hat{a}_{x/y}^\dag \ket{\alpha_i}_{x/y},
\end{equation}
where $\hat{a}_{x/y}^\dag$ is the creation operator of the photon field in the vicinity of $x$ and $y$, respectively.
This scenario can be formally described as
\begin{subequations}
\begin{eqnarray}
\ket{A}_x &=& \ket{\alpha_1}_x, \quad  \ket{B}_x = \ket{\alpha_1,1}_x, \\
\ket{A}_y &=& \ket{\alpha_2}_y, \quad  \ket{B}_y = \ket{\alpha_2,1}_y.
\end{eqnarray}
\end{subequations}
By \eqref{mu s}, we have
\begin{equation}
\mu_s = \abs{{}_x\!\bra{\alpha_1}\otimes{}_y\inner{\alpha_2,1}{\alpha_1,1}_x\otimes\ket{\alpha_2}_y}
=\frac{\abs{\alpha_1}\abs{\alpha_2}}{\sqrt{1+\abs{\alpha_1}^2}\sqrt{1+\abs{\alpha_2}^2}}.
\end{equation}
If $\abs{\alpha_1}\gg1$ and $\abs{\alpha_2}\gg1$, we have $\mu_s\approx1$, which implies $\mathcal{D}\approx0$ and $\mathcal{V}\approx\abs{\sin\theta_1}$ according to \eqref{D} and \eqref{V}. That is, for $\abs{\alpha_1}\gg1$ and $\abs{\alpha_2}\gg1$, the SPACSs $\ket{\alpha_1,1}_x$ and $\ket{\alpha_2,1}_y$ are almost indistinguishable from the coherent states $\ket{\alpha_1}_x$ and $\ket{\alpha_2}_y$, and therefore the which-way information is virtually unmarked and the interference pattern can yield nearly maximum visibility (by setting $\theta_1=\pm\pi/2$).
On the other hand, if $\abs{\alpha_1}$ or $\abs{\alpha_2}$ is close to $O(1)$, the purity $\mu_s$ is considerably smaller than unity, and correspondingly, to a certain degree, the which-way distinguishability is marked and the interference visibility is diminished.
The interferometer using the ENBS provides an appealing scenario whereby the which-way distinguishability and thus the interference visibility are controllable.\footnote{The complementarity relation between $\mathcal{D}$ and $\mathcal{V}$ is still given by \eqref{V D duality}. In \cite{sciadv.abi9268}, the complementarity of wave-particle duality is analyzed in detail, but it considers the distinguishability as defined in \cite{PhysRevResearch.2.012031}, instead of that defined in \eqref{D def}  (recall \ftref{foot:different D}).}

\subsection{Quantum erasure: optimal case}
Let us now study the Scully-Dr{\"u}hl-type quantum eraser as depicted in \figref{fig:SD interferometer}. In this subsection, we first consider the optimal case that no additional degrees of freedom are involved in the scenario of \figref{fig:transition} (c) or any equivalent scenarios.
Formally, the optimal case can be described as
\begin{subequations}\label{a ax}
\begin{eqnarray}
\ket{A}_x &=& \ket{a_1}_x\ket{0}_{\gamma_i}, \quad  \ket{B}_x = \ket{a_1}_x\ket{\Path{x}}_{\gamma_i}, \\
\ket{A}_y &=& \ket{a_2}_y\ket{0}_{\gamma_i}, \quad  \ket{B}_y = \ket{a_2}_y\ket{\Path{y}}_{\gamma_i},
\end{eqnarray}
\end{subequations}
where $\ket{\Path{x/y}}_{\gamma_i}$ represents the idler photon $\gamma_i$ on $\Path{x}$ or $\Path{y}$ and $\ket{0}_{\gamma_i}$ represents the vacuum state of $\gamma_i$.
Because $\inner{A_x}{B_x}=0$, $\inner{A_y}{B_y}=0$, and consequently $\inner{A_x,B_y}{B_xA_y}=0$, it follows from \eqref{SD interference} that $P(D_+)=P(D_-)=1/2$, showing no interference pattern in the total ensemble.

The entangled state $\ket{\Psi_0}$ in \eqref{Psi0} now reads as
\begin{equation}\label{Psi0 SD}
\ket{\Psi_0} =
\frac{1}{\sqrt{2}}\ket{a_1}_x\ket{a_2}_y
\left(\ket{\Path{1}}_{\gamma_s}\ket{\Path{x}}_{\gamma_i}
+ \ket{\Path{2}}_{\gamma_s}\ket{\Path{y}}_{\gamma_i}
\right).
\end{equation}
In comparison to \eqref{EPR pair}, we can make the analogy:
\begin{subequations}\label{analogy}
\begin{eqnarray}
\ket{\Path{1}}_{\gamma_s} & \longleftrightarrow & \ket{\leftrightarrow}_{\gamma_s},
\quad
\ket{\Path{2}}_{\gamma_s}  \longleftrightarrow  \ket{\updownarrow}_{\gamma_s},\\
\ket{\Path{x}}_{\gamma_i} & \longleftrightarrow & \ket{\updownarrow}_{\gamma_i},
\quad
\ket{\Path{y}}_{\gamma_i}  \longleftrightarrow  -\ket{\leftrightarrow}_{\gamma_i}.
\end{eqnarray}
\end{subequations}
Correspondingly, in accordance with \eqref{n + -}, we define
\begin{subequations}\label{n + - s}
\begin{eqnarray}
\ket{\hat{n}_{\vartheta,\varphi},+}_{\gamma_s} &:=& \cos(\vartheta/2)\ket{\Path{1}}_{\gamma_s} + e^{i\varphi} \sin(\vartheta/2)\ket{\Path{2}}_{\gamma_s},\\
\ket{\hat{n}_{\vartheta,\varphi},-}_{\gamma_s} &:=& \sin(\vartheta/2)\ket{\Path{1}}_{\gamma_s} - e^{i\varphi} \cos(\vartheta/2)\ket{\Path{2}}_{\gamma_s},
\end{eqnarray}
\end{subequations}
and
\begin{subequations}\label{n + - i}
\begin{eqnarray}
\ket{\hat{n}_{\vartheta,\varphi},+}_{\gamma_i} &:=& -\cos(\vartheta/2)\ket{\Path{y}}_{\gamma_i} + e^{i\varphi} \sin(\vartheta/2)\ket{\Path{x}}_{\gamma_i},\\
\ket{\hat{n}_{\vartheta,\varphi},-}_{\gamma_i} &:=& -\sin(\vartheta/2)\ket{\Path{y}}_{\gamma_i} - e^{i\varphi} \cos(\vartheta/2)\ket{\Path{x}}_{\gamma_i}.
\end{eqnarray}
\end{subequations}
The entangled state $\ket{\Psi_0}$ then can be recast as
\begin{equation}\label{Psi0 SD 2}
\ket{\Psi_0} = -\frac{e^{i\varphi}}{\sqrt{2}}\ket{a_1}_x\ket{a_2}_y \otimes
\left(\ket{\hat{n}_{\vartheta,\varphi},+}_{\gamma_s}\ket{\hat{n}_{\vartheta,\varphi},-}_{\gamma_i}
-\ket{\hat{n}_{\vartheta,\varphi},-}_{\gamma_s}\ket{\hat{n}_{\vartheta,\varphi},+}_{\gamma_i}\right),
\end{equation}
of which the ``spinor'' part bears exactly the same mathematical structure as that in \eqref{EPR pair 2}.
Accordingly to \eqref{evolution path}, the state $\ket{\hat{n}_{\vartheta,\varphi},\pm}_{\gamma_s}$ undergoes the evolution
\begin{subequations}\label{evolution gamma s}
\begin{eqnarray}
\ket{\hat{n}_{\vartheta,\varphi},+}_{\gamma_s}
& \longrightarrow &
\left(\alpha_s\cos(\vartheta/2)e^{i\phi_1} + \beta_s\sin(\vartheta/2)e^{i\varphi}\right)\ket{D_+}_{\gamma_s}
\nonumber\\
&& \quad \mbox{} +
\left(\beta_s^*\cos(\vartheta/2)e^{i\phi_1} - \alpha_s^*\sin(\vartheta/2)e^{i\varphi}\right)\ket{D_-}_{\gamma_s}, \nonumber\\
&\equiv&
\inner{\hat{n}_{\theta_1,\phi_1},+}{\hat{n}_{\vartheta,\varphi},+}\ket{D_+}_{\gamma_s}
+
\inner{\hat{n}_{\theta_1,\phi_1},-}{\hat{n}_{\vartheta,\varphi},+}\ket{D_-}_{\gamma_s}, \\
\ket{\hat{n}_{\vartheta,\varphi},-}_{\gamma_s}
& \longrightarrow &
\left(\alpha_s\sin(\vartheta/2)e^{i\phi_1} - \beta_s\cos(\vartheta/2)e^{i\varphi}\right)\ket{D_+}_{\gamma_s}
\nonumber\\
&& \quad \mbox{} +
\left(\beta_s^*\sin(\vartheta/2)e^{i\phi_1} + \alpha_s^*\cos(\vartheta/2)e^{i\varphi}\right)\ket{D_-}_{\gamma_s}, \nonumber\\
&\equiv&
e^{i\phi}
\inner{\hat{n}_{\theta_1,\phi_1},+}{\hat{n}_{\vartheta,\varphi},-}\ket{D_+}_{\gamma_s}
+
\inner{\hat{n}_{\theta_1,\phi_1},-}{\hat{n}_{\vartheta,\varphi},-}\ket{D_-}_{\gamma_s},
\end{eqnarray}
\end{subequations}
where $\alpha_s=\cos(\theta_1/2)$ and $\beta_s=\sin(\theta_1/2)$ are the coefficients of the beam splitter $\mathrm{BS}_s$ and $\phi_1$ is the phase shift along $\Path{1}$.
Similarly, the state $\ket{\hat{n}_{\vartheta,\varphi},\pm}_{\gamma_i}$ undergoes the evolution
\begin{subequations}\label{evolution gamma i}
\begin{eqnarray}
\ket{\hat{n}_{\vartheta,\varphi},+}_{\gamma_i}
& \longrightarrow &
\left(\alpha_i\cos(\vartheta/2)e^{i\phi_2} + \beta_i\sin(\vartheta/2)e^{i\varphi}\right)\ket{D'_+}_{\gamma_i}
\nonumber\\
&& \quad \mbox{} +
\left(\beta_i^*\cos(\vartheta/2)e^{i\phi_2} - \alpha_i^*\sin(\vartheta/2)e^{i\varphi}\right)\ket{D'_-}_{\gamma_i}, \nonumber\\
&\equiv&
\inner{\hat{n}_{\theta_2,\phi_2},+}{\hat{n}_{\vartheta,\varphi},+}\ket{D'_+}_{\gamma_i}
+
\inner{\hat{n}_{\theta_2,\phi_2},-}{\hat{n}_{\vartheta,\varphi},+}\ket{D'_-}_{\gamma_i}, \\
\ket{\hat{n}_{\vartheta,\varphi},-}_{\gamma_i}
& \longrightarrow &
\left(\alpha_i\sin(\vartheta/2)e^{i\phi_2} - \beta_i\cos(\vartheta/2)e^{i\varphi}\right)\ket{D'_+}_{\gamma_i}
\nonumber\\
&& \quad \mbox{} +
\left(\beta_i^*\sin(\vartheta/2)e^{i\phi_2} + \alpha_i^*\cos(\vartheta/2)e^{i\varphi}\right)\ket{D'_-}_{\gamma_i}, \nonumber\\
&\equiv&
e^{i\phi}
\inner{\hat{n}_{\theta_2,\phi_2},+}{\hat{n}_{\vartheta,\varphi},-}\ket{D'_+}
+
\inner{\hat{n}_{\theta_2,\phi_2},-}{\hat{n}_{\vartheta,\varphi},-}\ket{D'_-},
\end{eqnarray}
\end{subequations}
where $\alpha_i=\cos(\theta_2/2)$ and $\beta_i=\sin(\theta_2/2)$ are the coefficients of the beam splitter $\mathrm{BS}_i$ and $\phi_2$ is the phase shift along $\Path{y}$.
The equations \eqref{evolution gamma s} and \eqref{evolution gamma i} are exactly analogous to \eqref{evolution}.

Consequently, by the analogy \eqref{analogy}, the setup of \figref{fig:SD interferometer} shares exactly the same mathematical structure as the setup of \figref{fig:two interferometers} insofar as the outcomes of $D_\pm$ and $D'_\pm$ and their correlation between them are concerned.
Therefore, the entanglement quantum eraser depicted in \figref{fig:two interferometers} and the Scully-Dr{\"u}hl-type quantum eraser depicted in \figref{fig:SD interferometer} are formally equivalent to each other and therefore have the same quantum erasure behavior as described by \eqref{prob sub}.

Conceptually, however, the Scully-Dr{\"u}hl-type quantum eraser bears some crucial difference from the entanglement quantum eraser.
As has been discussed in \secref{sec:MZ interferometer}, each of the two Mach-Zehnder interferometers in \figref{fig:two interferometers} functions as a projective measurement in the same way as the Stern-Gerlach apparatus does. This is not the case for the two Mach-Zehnder interferometers in \figref{fig:SD interferometer}.
Consider a signal photon produced in the state $\ket{\Path{1}}_{\gamma_s}$ at $x$. It will travel along $\Path{1}$ and eventually click $D_+$ with probability $\cos^2(\theta_1/2)$ or $D_-$ with probability $\sin^2(\theta_1/2)$. These probabilities are simply determined by the transmission-reflection parameter $\theta_1$ of the beam splitter $\mathrm{BS}_s$.
On the other hand, in analogy to \eqref{special case}, the state $\ket{\Path{1}}_{\gamma_s}$ can be recast as
\begin{equation}
\ket{\Path{1}}_{\gamma_s} = \cos(\theta_1/2)\ket{\hat{n}_{\theta_1,\phi_1},+}_{\gamma_s} + \sin(\theta_1/2)\ket{\hat{n}_{\theta_1,\phi_1},-}_{\gamma_s},
\end{equation}
by which it is tempting to assert that the Mach-Zehnder interferometer parameterized by $\theta_1$ and $\phi_1$ as a whole projects the state $\ket{\Path{1}}_{\gamma_s}$ into $\ket{\hat{n}_{\theta_1,\phi_1},+}_{\gamma_s}$ or $\ket{\hat{n}_{\theta_1,\phi_1},-}_{\gamma_s}$.
However, this assertion is conceptually problematic against semiclassical reasoning, although it is mathematically legitimate. According to \eqref{n + - s}, the states $\ket{\hat{n}_{\theta_1,\phi_1},+}_{\gamma_s}$ and $\ket{\hat{n}_{\theta_1,\phi_1},-}_{\gamma_s}$ have both parts of $\ket{\Path{1}}_{\gamma_s}$ and $\ket{\Path{2}}_{\gamma_s}$ unless $\theta_1=0$ or $\theta_1=\pi$. Thus, saying that the state $\ket{\Path{1}}_{\gamma_s}$, which by definition travels \emph{only} $\Path{1}$, can be projected into a superposition of both $\ket{\Path{1}}_{\gamma_s}$ and $\ket{\Path{2}}_{\gamma_s}$ is conceptually self-contradictory. A more sensible interpretation is that the projection is made by the beam splitter $\mathrm{BS}_s$, not the Mach-Zehnder interferometer \emph{as a whole}.\footnote{By contrast, in \eqref{special case} for the Mach-Zehnder interferometer in \figref{fig:interferometer}, asserting that the state $\ket{\leftrightarrow}$ is projected into $\ket{\hat{n}_{\theta,\phi},+}:=\cos(\theta/2)\ket{\leftrightarrow} + e^{i\phi} \sin(\theta/2)\ket{\updownarrow}$ or $\ket{\hat{n}_{\theta,\phi},-}:=\sin(\theta/2)\ket{\leftrightarrow} - e^{i\phi} \cos(\theta/2)\ket{\updownarrow}$ is perhaps counterintuitive, but not self-contradictory. Recall the discussion after \eqref{special case}.}
Therefore, although the Scully-Dr{\"u}hl-type quantum eraser in \figref{fig:SD interferometer} is formally equivalent to the entanglement quantum eraser in \figref{fig:two interferometers}, the former is conceptually more different from the EPR-Bohm experiment in \figref{fig:EPR} than the latter is.

Furthermore, in the entanglement quantum eraser as shown in \figref{fig:two interferometers}, the which-way information of the signal photon $\gamma_s$ becomes factual only when $\theta_1=0,\pi$, whereas the which-way information of the idler photon $\gamma_i$ becomes factual only when $\theta_2=0,\pi$.
By contrast, in the Scully-Dr{\"u}hl-type quantum eraser as shown in \figref{fig:SD interferometer}, the which-way information of $\gamma_s$ becomes factual when $\theta_1=0,\pi$ or $\theta_2=0,\pi$, and the which-way information of $\gamma_i$ becomes factual also when $\theta_1=0,\pi$ or $\theta_2=0,\pi$.
Accordingly, in \figref{fig:two interferometers}, the outcome of $D'_\pm$ measured by Bob always yields a counterfactual conviction about whether $\gamma_s$ has traveled or will travel along $\Path{1}$, $\Path{2}$, or both. On the other hand, in \figref{fig:SD interferometer}, if Bob chooses $\theta_2=0,\pi$, the outcome of $D'_\pm$ yields a factual description about whether $\gamma_s$ has traveled or will travel along either $\Path{1}$ or $\Path{2}$.
This subtle difference raises a profound conceptual issue.
As has been discussed in \secref{sec:comparison to EPR}, for the entanglement quantum eraser in \figref{fig:two interferometers}, whether $\gamma_s$ travels $\Path{1}$, $\Path{2}$, or both can be influenced by the choice made by Bob, but this interpretation makes sense only if one employs the counterfactual reasoning about which-way information. If one is willing to dismiss all counterfactual reasoning, the entanglement quantum eraser does not present any additional mystery beyond the standard EPR puzzle.
By contrast, for the Scully-Dr{\"u}hl-type quantum eraser in \figref{fig:SD interferometer}, whether the way $\gamma_s$ travels can be influenced by the choice made by Bob is no longer a matter of purely counterfactual reasoning, as indeed the which-way information of $\gamma_s$ can be made factual by Bob's choice. In this sense, the Scully-Dr{\"u}hl-type quantum eraser presents a conceptual ``mystery'' beyond the standard EPR puzzle, even if one dismisses all counterfactual reasoning.

To underscore the ``mystery'' further, let us consider the special case that Alice sets $\theta_1\neq0,\pi$ and performs her measurement first, while Bob sets $\theta_2=0$ and performs his measurement in a delayed-choice manner. According to \eqref{prob sub}, in this case we have
\begin{subequations}\label{prob sub special condition}
\begin{eqnarray}
P(D_\pm|D'_+) &=& \frac{1\mp \hat{n}_{\theta_1,\phi_1}\cdot\hat{z}}{2}
=
\left\{\begin{array}{c}
         \sin^2(\theta_1/2) \\
         \cos^2(\theta_1/2)
       \end{array}
\right.,\\
P(D_\pm|D'_-) &=& \frac{1\pm \hat{n}_{\theta_1,\phi_1}\cdot\hat{z}}{2}
=
\left\{\begin{array}{c}
         \cos^2(\theta_1/2) \\
         \sin^2(\theta_1/2)
       \end{array}
\right..
\end{eqnarray}
\end{subequations}
When Alice obtains a signal in $D_\pm$, from her standpoint, the entangled state $\ket{\Psi_0}$ in \eqref{Psi0 SD 2} is collapsed into
\begin{equation}
\ket{\hat{n}_{\theta_1,\phi_1},\pm}_{\gamma_s}\ket{\hat{n}_{\theta_1,\phi_1},\mp}_{\gamma_i}.
\end{equation}
By \eqref{n + - s}, the state $\ket{\hat{n}_{\theta_1,\phi_1},\pm}_{\gamma_s}$ contains both parts of $\ket{\Path{1}}_{\gamma_s}$ and $\ket{\Path{2}}_{\gamma_s}$ since $\theta_1\neq0,\pi$.
Therefore, Alice preferably (but not logically necessarily) deduces the which-way history of $\gamma_s$ to be having traveled \emph{both} $\Path{1}$ and $\Path{2}$. Note that the which-way history of $\gamma_s$ deduced by Alice is \emph{counterfactual} since $\theta_1\neq0,\pi$.
Later, Bob performs his measurement and obtains a signal in $D'_\pm$. From his standpoint, the entangled state $\ket{\Psi_0}$ is collapsed into
\begin{equation}
\ket{\hat{n}_{\theta_2=0,\phi_2},\mp}_{\gamma_s}\ket{\hat{n}_{\theta_2=0,\phi_2},\pm}_{\gamma_i}
\equiv
\ket{\hat{z},\mp}_{\gamma_s}\ket{\hat{z},\pm}_{\gamma_i}
\equiv
\left\{
\begin{array}{c}
  \ket{\Path{2}}_{\gamma_s}\ket{\Path{y}}_{\gamma_i} \\
  -\ket{\Path{1}}_{\gamma_s}\ket{\Path{x}}_{\gamma_i}
\end{array}
\right.,
\end{equation}
where \eqref{n + - s} and \eqref{n + - i} have been used. After the signal is registered in $D'_\pm$, there is no ambiguity any more about how $\gamma_s$ and $\gamma_i$ have traveled. Therefore, Bob \emph{logically necessarily} deduces the which-way history of $\gamma_s$ to be having traveled \emph{only} $\Path{2}$ or \emph{only} $\Path{1}$ and the which-way history of $\gamma_i$ to be having traveled \emph{only} $\Path{y}$ or \emph{only} $\Path{x}$ correspondingly, depending on whether the signal is registered in $D'_+$ or $D'_-$, respectively. Note that the which-way histories of $\gamma_s$ and $\gamma_i$ deduced by Bob are \emph{factual} since $\theta_2=0$.
Therefore, after the signal in $D'_\pm$ is obtained, the \emph{counterfactual} history of $\gamma_s$ deduced by Alice is ``overridden'' by the \emph{factual} history of $\gamma_s$ deduced by Bob. In this sense, it is said that the history of $\gamma_s$ can be affected by the choice of Bob, even retroactively.\footnote{This does not violate causality, because the overriding resets a counterfactual history into a factual one, not a factual history into a different factual one. In the case that Alice sets $\theta_1=0,\pi$ and Bob sets $\theta_2=0,\pi$, the histories deduced by Alice and Bob both are factual, and they just agree with each other --- no overriding upon each other.}
In view of this factual history, each individual $\gamma_s$ in the subensemble associated with $D'_+$ travels only $\Path{2}$ and each $\gamma_s$ in the subensemble associated with $D'_-$ travels only $\Path{1}$. Correspondingly, the conditional probabilities in \eqref{prob sub special condition} are said to be caused solely by the transmission-reflection parameter $\theta_1$ of the beam splitter $\mathrm{BS}_s$.

What is really puzzling is that we are not allowed to simply declare that the counterfactual history deduced by Alice is meaningless and has to be ``factualized'' sooner or later. If Alice sets $\theta_1\neq0,\pi$ and Bob also sets $\theta_2\neq0,\pi$, the conditional probabilities \eqref{prob sub} will exhibit the two-path interference pattern. Therefore, the counterfactual history that $\gamma_s$ travels both $\Path{1}$ and $\Path{2}$ makes good sense and is not to be factualized. The ``mystery'' of the Scully-Dr{\"u}hl-type quantum eraser beyond the standard EPR puzzle is that whether the counterfactual history of $\gamma_s$ is to be factualized or not really depends on the choice made by Bob. The EPR-Bohm experiment does not have this issue.

\subsection{Quantum erasure: nonoptimal case}
In this subsection, we study the Scully-Dr{\"u}hl-type quantum eraser in the nonoptimal case in which remaining degrees of freedom of the atoms as well as external systems and fields are involved. To take into account additional degrees of freedom, similar to \eqref{extension 2}, we now employ the substitution:
\begin{subequations}\label{extension 3}
\begin{eqnarray}
\ket{A_x,B_y} &\dashrightarrow& \ket{a_1}_x\ket{a_2}_y\ket{\Path{y}}_{\gamma_i}\otimes\ket{m}, \\
\ket{B_x,A_y} &\dashrightarrow& \ket{a_1}_x\ket{a_2}_y\ket{\Path{x}}_{\gamma_i}\otimes\ket{n},
\end{eqnarray}
\end{subequations}
where $\ket{m}$ and $\ket{n}$ are given in \eqref{m n}.
The entangled state \eqref{Psi0 SD} is then modified into
\begin{equation}
\ket{\Psi_0} =
\frac{1}{\sqrt{2}}\ket{a_1}_x\ket{a_2}_y
\left(\ket{+}_{\gamma_s}\ket{-}_{\gamma_i}\ket{m}
- \ket{-}_{\gamma_s}\ket{+}_{\gamma_i}\ket{n}
\right),
\end{equation}
where we have adopted the shorthand notations $\ket{+}:=(1,0)^\mathrm{T}$ and $\ket{-}:=(0,1)^\mathrm{T}$ in accordance with the doublet representation in \eqref{n + -}, i.e.,
\begin{subequations}\label{+ - Path}
\begin{eqnarray}
\ket{+}_{\gamma_s} &=& \ket{\Path{1}}_{\gamma_s}, \qquad \ket{-}_{\gamma_s} = \ket{\Path{2}}_{\gamma_s}, \\
\ket{+}_{\gamma_i} &=& -\ket{\Path{y}}_{\gamma_i}, \qquad \ket{-}_{\gamma_i} = \ket{\Path{x}}_{\gamma_s}.
\end{eqnarray}
\end{subequations}
Because $\inner{A_x,B_y}{B_xA_y}=0$, it follows from \eqref{SD interference} that $P(D_+)=P(D_-)=1/2$, showing no interference pattern in the total ensemble.

When $\gamma_i$ clicks $D'_+$ or $D'_-$, from Bob's standpoint, $\ket{\Psi_0}$ is collapsed into
\begin{equation}
\ket{\Psi_{D'_\pm}} =
\ket{a_1}_x\ket{a_2}_y \ket{\hat{n}_{\theta_2,\phi_2},\pm}_{\gamma_i}
\otimes
\left(
\zeta\ket{+}_{\gamma_s}\ket{m}
-
\eta\ket{-}_{\gamma_s}\ket{n}
\right),
\end{equation}
where
\begin{subequations}
\begin{eqnarray}
\zeta &:=& \inner{\hat{n}_{\theta_2,\phi_2},\pm}{-},\\
\eta &:=& \inner{\hat{n}_{\theta_2,\phi_2},\pm}{+}.
\end{eqnarray}
\end{subequations}
The reduced density matrix for $\gamma_s$ associated with $\ket{\Psi_{D'_\pm}}$ is given by tracing out all degrees of freedom other than $\gamma_s$, i.e.,
\begin{eqnarray}
\rho_{D'_\pm}^{(\gamma_s)} &=& \Tr_{x\otimes y\otimes\gamma_i\otimes m/n} \ket{\Psi_{D'_\pm}}\bra{\Psi_{D'_\pm}} \nonumber\\
&=&
\abs{\zeta}^2\ket{+}\bra{+}
+\abs{\eta}^2\ket{-}\bra{-}
-\zeta\eta^*\inner{n}{m}\ket{+}\bra{-}
-\zeta^*\eta\inner{m}{n}\ket{-}\bra{-}.
\end{eqnarray}
Consequently, within the confines of the subensemble associated with $D'_+$ or $D'_-$, the detection probabilities of $D_+$ and $D_-$ are given by
\begin{eqnarray}\label{prob sub nonoptimal}
P(D_\pm|D'_{\pm'}) &=& \Tr\left( \ket{\hat{n}_{\theta_1,\phi_1}} \bra{\hat{n}_{\theta_1,\phi_1}} \rho_{D'_\pm}^{(\gamma_s)} \right) \nonumber\\
&=& \abs{\inner{\hat{n}_{\theta_2,\phi_2},\mp'}{+}\inner{+}{\hat{n}_{\theta_1,\phi_1},\pm}\ket{n}
+\inner{\hat{n}_{\theta_2,\phi_2},\mp'}{-}\inner{-}{\hat{n}_{\theta_1,\phi_1},\pm}\ket{m}}^2,
\end{eqnarray}
where we have used $\zeta\eta^*\equiv \inner{\hat{n}_{\theta_2,\phi_2},\pm}{-} \inner{+}{\hat{n}_{\theta_2,\phi_2},\pm} = -\inner{\hat{n}_{\theta_2,\phi_2},\mp}{-} \inner{+}{\hat{n}_{\theta_2,\phi_2},\mp}$ by \eqref{n + -}.

If the process of producing $\gamma_s$ and $\gamma_i$ does not leave any footprint upon external systems or the remaining degrees of freedom of the atoms, we have $\ket{m}=\ket{n}$. In this case, \eqref{prob sub nonoptimal} obviously is reduced to the optimal result \eqref{prob sub}.
Otherwise, computing \eqref{prob sub nonoptimal} explicitly leads to
\begin{subequations}\label{prob sub nonoptimal 2}
\begin{eqnarray}
P(D_+|D'_+) &=& \cos^2\frac{\theta_1}{2}\sin^2\frac{\theta_2}{2} + \sin^2\frac{\theta_1}{2}\cos^2\frac{\theta_2}{2}
-\frac{\mu_s}{2}\sin\theta_1\sin\theta_2\cos(\phi_1-\phi_2-\delta), \\
P(D_-|D'_+) &=& \sin^2\frac{\theta_1}{2}\sin^2\frac{\theta_2}{2} + \cos^2\frac{\theta_1}{2}\cos^2\frac{\theta_2}{2}
+\frac{\mu_s}{2}\sin\theta_1\sin\theta_2\cos(\phi_1-\phi_2-\delta), \\
P(D_+|D'_-) &=& \cos^2\frac{\theta_1}{2}\cos^2\frac{\theta_2}{2} + \sin^2\frac{\theta_1}{2}\sin^2\frac{\theta_2}{2}
+\frac{\mu_s}{2}\sin\theta_1\sin\theta_2\cos(\phi_1-\phi_2-\delta), \\
P(D_-|D'_-) &=& \sin^2\frac{\theta_1}{2}\cos^2\frac{\theta_2}{2} + \cos^2\frac{\theta_1}{2}\sin^2\frac{\theta_2}{2}
-\frac{\mu_s}{2}\sin\theta_1\sin\theta_2\cos(\phi_1-\phi_2-\delta),
\end{eqnarray}
\end{subequations}
where \eqref{m n product} has been used.
Therefore, within each subensemble associated with $D'_+$ or $D'_-$, the interference pattern as modulated in response to $\phi_1$ can still be recovered but only to the extent diminished by a factor of $\mu_s$ compared to the optimal result.\footnote{In a real experiment, the interference pattern recovered by the quantum erasure is further diminished much more than \eqref{prob sub nonoptimal 2} because of the reason addressed in \ftref{foot:real experiment}.}
In the case that $\ket{m}$ is completely different from $\ket{n}$, we have $\inner{m}{n}=0$ and consequently the interference pattern can never be recovered by the quantum erasure effect.
The analysis presented in this subsection also applies to the entanglement quantum eraser as shown in \figref{fig:two interferometers}, if the entangled signal-idler photon pair involves additional degrees of freedom.

In the double-slit experiment of the Scully-Dr{\"u}hl-type quantum eraser performed by Kim \textit{et al.}\ \cite{PhysRevLett.84.1}, instead of two atomic sources, a pump laser beam is divided by a double slit and directed into two regions inside a BBO crystal to produce a signal-idler photon pair by SPDC. As the SPDC slits have a finite width, they introduce additional degrees of freedom other than two ideal point sources \cite{PhysRevResearch.2.012031}. As a result of the additional degrees of freedom due to the finite width, the visibility of the interference pattern recovered by the quantum erasure is considerably diminished compared to the case of two ideal point slits (particularly, there are no completely destructive fringes).
Similarly, in the double-slit experiment of the entanglement quantum eraser performed by Walborn \textit{et al.}\ \cite{PhysRevA.65.033818}, the two slits have a finite width and as a result the interference pattern recovered by the quantum erasure is also considerably diminished.

\subsection{Recording and erasing of which-way information}
We have looked into the similarities and dissimilarities between the entanglement quantum eraser and the Scully-Dr{\"u}hl-type quantum eraser.

In the entanglement quantum eraser as depicted in \figref{fig:two interferometers}, the polarizing beam splitter $\mathrm{PBS}$ is used in \figref{fig:interferometer} to couple the which-way information of a photon with its polarization. Therefore, in a sense, the which-way information of the signal photon $\gamma_s$ is ``recorded'' in terms of the polarization of the idler photon $\gamma_i$.
In other kinds of entanglement quantum erasers, likewise, the which-way information of the signal quanton (also designated as $\gamma_s$) is recorded in terms of some \emph{internal state} of the idler quanton (also designated as $\gamma_i$). The internal state of $\gamma_i$ can be measured in different ways by the choice of Bob to either read out or erase the recorded information.
However, the which-way information of $\gamma_s$ inferred from the internal state of $\gamma_i$ is always counterfactual, as the inference is based on the counterfactual reasoning. In this sense, the internal state of $\gamma_i$ does not really serves as a ``recorder'' of the which-way information of $\gamma_s$.
Accordingly, if all counterfactual reasoning is dismissed, one can maintain that the which-way information of $\gamma_s$ is not recorded in the first place at all, let alone erased in a later time. From this perspective, the entanglement quantum eraser does not present any additional mystery beyond the standard EPR-Bohm experiment, as advocated in \cite{kastner2019delayed, qureshi2020}.

By contrast, in the Scully-Dr{\"u}hl-type quantum eraser as depicted in \figref{fig:SD interferometer}, the polarization (or some internal state) is not coupled with which-way information, and the internal state of $\gamma_i$ is not used to deduce the which-way information of $\gamma_s$. Instead, the which-way information of $\gamma_s$ is ``recorded'' in terms of the states of the two objects at $x$ and $y$. Because the two objects are \emph{spatially separated}, the which-way information of $\gamma_s$ inferred from the measurement upon the states of two objects becomes factual if the measurement, such as the UQSD given by \eqref{Fs}, yields a conclusive outcome. That is, the counterfactual history of $\gamma_s$ is ``factualized'' by the conclusive outcome. Therefore, it makes good sense to say that the two objects serve as the ``recorders'' of the which-way information of $\gamma_s$. Correspondingly, if the potentiality to yield a conclusive outcome is eliminated by the choice made by Bob, it makes good sense to say that the record of the which-way information of $\gamma_s$ is ``erased''. In other words, the potentiality to factualize the counterfactual history of $\gamma_s$ is eliminated.

Although \emph{formally} equivalent to the EPR-Bohm experiment, the entanglement quantum eraser is \emph{conceptually} quite different from it, and the Scully-Dr{\"u}hl-type quantum eraser is even more so.
As opposed to what advocated in \cite{kastner2019delayed, qureshi2020}, the ``mystery'' of quantum erasure cannot be completely relegated to nothing but the EPR correlation.
Furthermore, the notion of ``quantum erasure'' may be inappropriate and misleading for the entanglement quantum eraser as suggested in \cite{kastner2019delayed, qureshi2020}, but it makes good sense for the Scully-Dr{\"u}hl-type quantum eraser.

\section{Many-world interpretation}\label{sec:MWI}
So far, we have considered the conceptual issues of the quantum eraser in the standard (i.e., Copenhagen) interpretation of quantum mechanics. In this section, we reconsider them in the many-worlds interpretation (MWI) \cite{RevModPhys.29.454, dewitt1975many, RevModPhys.29.463}. From the practical point of view, the choice between the Copenhagen interpretation and the MWI is only a matter of taste, as they make identical experimental predictions (except, perhaps, the grotesque experiment of ``quantum suicide'') \cite{tegmark1998}. Nevertheless, the MWI is said to resolve the measurement problem and many other paradoxes of quantum mechanics (including the quantum eraser) in a simple way, since it does not invoke the concept of wavefunction collapse as all possible experimental outcomes exist simultaneously in the universal wavefunction \cite{dewitt1970}.
The aim of this section is not to support or refute the MWI, but to make clear its conceptual merits and demerits in view of the peculiar setting of the quantum eraser, particularly of the Scully-Dr{\"u}hl type.

In the MWI formalism, the combined system of the measured objects and the measuring apparatuses, as an isolated system as a whole, is described by a universal wavefunction, which evolves unitarily according to the Schr\"{o}dinger equation.
For the Scully-Dr{\"u}hl-type quantum eraser depicted in \figref{fig:SD interferometer}, the initial state of the universal function is given by \eqref{Psi0 SD} augmented with the degrees of the detectors $D_\pm$ and $D'_\pm$. That is, ignoring the part $\ket{a_1}_x\ket{a_2}_y$ for simplicity as it does not evolve any more and adopting the notation \eqref{+ - Path}, we have
\begin{equation}\label{Psi0 MWI}
\ket{\Psi_0} =
\frac{1}{\sqrt{2}}
\left(
\ket{+}_{\gamma_s}\ket{-}_{\gamma_i}
-\ket{-}_{\gamma_s}\ket{+}_{\gamma_i}\right)
\otimes \ket{\tilde{D}}\ket{\tilde{D}'},
\end{equation}
where $\ket{\tilde{D}}$ represents the state of the combined system of $D_+$ and $D_-$ with no signal registered, and $\ket{\tilde{D}'}$ represents the state of the combined system of $D'_+$ and $D'_-$ with no signal registered.
As $\gamma_s$ enters the interferometer parameterized by $\theta_1$ and $\phi_1$ and $\gamma_i$ enters the interferometer parameterized by $\theta_2$ and $\phi_2$, the state $\ket{\Psi_0}$ undergoes the evolution govern by the same form of \eqref{evolution} (except that the degree of polarization is not considered). Accordingly, the state $\ket{\Psi_0}$ evolves into the new state
\begin{eqnarray}\label{Psi MWI}
\ket{\Psi}
&=& \frac{1}{\sqrt{2}}
\Big(
\left(
\inner{\hat{n}_{\theta_1,\phi_1},+}{+}\ket{D_+}_{\gamma_s}
+
\inner{\hat{n}_{\theta_1,\phi_1},-}{+}\ket{D_-}_{\gamma_s}
\right) \nonumber\\
&&\qquad\quad \mbox{}\times
\left(
\inner{\hat{n}_{\theta_2,\phi_2},+}{-}\ket{D_+}_{\gamma_i}
+
\inner{\hat{n}_{\theta_2,\phi_2},-}{-}\ket{D_-}_{\gamma_i}
\right) \nonumber\\
&&\qquad \mbox{}-
\left(
\inner{\hat{n}_{\theta_1,\phi_1},+}{-}\ket{D_+}_{\gamma_s}
+
\inner{\hat{n}_{\theta_1,\phi_1},-}{-}\ket{D_-}_{\gamma_s}
\right) \nonumber\\
&&\qquad\quad \mbox{}\times
\left(
\inner{\hat{n}_{\theta_2,\phi_2},+}{+}\ket{D_+}_{\gamma_i}
+
\inner{\hat{n}_{\theta_2,\phi_2},-}{+}\ket{D_-}_{\gamma_i}
\right)
\Big)
\otimes \ket{\tilde{D}}\ket{\tilde{D}'},
\end{eqnarray}
where $\ket{D_\pm}_{\gamma_s}$ represents the $\gamma_s$ state that is going to strike $D_\pm$, and $\ket{D_\pm}_{\gamma_i}$ represents the $\gamma_i$ state that is going to strike $D'_\pm$.
Proceeding to the next step, the evolution due to the interaction between the photons and the detectors is given by
\begin{subequations}\label{D evolution}
\begin{eqnarray}
\ket{D_\pm}_{\gamma_s}\ket{\tilde{D}}&\longrightarrow&\ket{\tilde{D}_\pm},\\
\ket{D_\pm}_{\gamma_i}\ket{\tilde{D}}&\longrightarrow&\ket{\tilde{D}'_\pm},
\end{eqnarray}
\end{subequations}
where $\ket{\tilde{D}_\pm}$ represents the state of the combined system of $D_+$ and $D_-$ with a signal registered in $D_+$ or $D_-$, and $\ket{\tilde{D}'_\pm}$ represents the state of the combined system of $D'_+$ and $D'_-$ with a signal registered in $D'_+$ or $D'_-$.
Computing \eqref{Psi MWI} explicitly with \eqref{D evolution}, we obtain the final state of the universal wavefunction as
\begin{eqnarray}\label{psi f MWI}
\ket{\Psi_f}&=& \frac{e^{-i\phi_2}}{\sqrt{2}}
\left(
\inner{\hat{n}_{\theta_1,\phi_1},+}{\hat{n}_{\theta_2,\phi_2},-}\ket{\tilde{D}_+}\ket{\tilde{D}'_+}
-\inner{\hat{n}_{\theta_1,\phi_1},+}{\hat{n}_{\theta_2,\phi_2},+}\ket{\tilde{D}_+}\ket{\tilde{D}'_-}
\right. \\
&&\qquad
\left.
\mbox{} +
\inner{\hat{n}_{\theta_1,\phi_1},-}{\hat{n}_{\theta_2,\phi_2},-}\ket{\tilde{D}_-}\ket{\tilde{D}'_+}
-\inner{\hat{n}_{\theta_1,\phi_1},-}{\hat{n}_{\theta_2,\phi_2},+}\ket{\tilde{D}_-}\ket{\tilde{D}'_-}
\right). \nonumber
\end{eqnarray}
The universal wavefunction ``branches'' into four different states, each of which represents a different measurement outcome $(D_+,D'_+)$, $(D_+,D'_-)$, $(D_-,D'_+)$, or $(D_-,D'_-)$. The different measurement outcomes exist simultaneously in the superposition. The \emph{relative frequencies} of the different possible outcomes are given by the squared norms of the corresponding coefficients, i.e.,
\begin{subequations}\label{prob MWI}
\begin{eqnarray}
P(D_+,D'_+) &=& \frac{1}{2}\abs{\inner{\hat{n}_{\theta_1,\phi_1},+}{\hat{n}_{\theta_2,\phi_2},-}}^2, \\
P(D_+,D'_-) &=& \frac{1}{2}\abs{\inner{\hat{n}_{\theta_1,\phi_1},+}{\hat{n}_{\theta_2,\phi_2},+}}^2, \\
P(D_-,D'_+) &=& \frac{1}{2}\abs{\inner{\hat{n}_{\theta_1,\phi_1},-}{\hat{n}_{\theta_2,\phi_2},-}}^2, \\
P(D_-,D'_-) &=& \frac{1}{2}\abs{\inner{\hat{n}_{\theta_1,\phi_1},-}{\hat{n}_{\theta_2,\phi_2},+}}^2,
\end{eqnarray}
\end{subequations}
which obviously gives rise to the same marginal probabilities as \eqref{flat prob} and the same conditional probabilities as \eqref{prob sub}.

In regard to the prediction of measurement outcomes, the Copenhagen interpretation and the MWI really make no difference. From the ``shut-up-and-calculate'' perspective, they just prescribe two different schemes of bookkeeping for performing calculation.
In the Copenhagen interpretation, different measurement outcomes are calculated separately; when a particular outcome is considered, the irrelevant parts are immediately discarded in the bookkeeping (i.e., the superposition is ``collapsed'' into the measured eigenspace).
By contrast, in the MWI, all possible outcomes are always kept and calculated altogether in the all-encompassing bookkeeping (i.e, the universal wavefunction).
The comparison between the calculation toward \eqref{flat prob} and \eqref{prob sub} in the Copenhagen interpretation and the calculation toward \eqnref{prob MWI} in the MWI demonstrates this point beyond a trivial model.

Conceptually, however, the Copenhagen interpretation and the MWI provide two very different ontological pictures. In the Copenhagen interpretation, as we have argued, whether $\gamma_s$ travels $\Path{1}$, $\Path{2}$, or both and whether $\gamma_i$ travels $\Path{x}$, $\Path{y}$, or both can be affected, even retroactively, by the choice of $\theta_1$ made by Alice and/or the choice of $\theta_2$ made by Bob.
By contrast, in the MWI, $\gamma_s$ and $\gamma_i$ always travel through their both paths, as $\ket{+}_{\gamma_s}$, $\ket{-}_{\gamma_s}$, $\ket{+}_{\gamma_i}$, and $\ket{-}_{\gamma_i}$ all persist from \eqref{Psi0 MWI} to \eqref{Psi MWI}. In the MWI, the ``mystery'' of the quantum eraser is simply resolved, since there is no such issue whether a quanton in a two-path interferometer travels either of the two paths or both of them --- it unambiguously travels through both \cite{carroll2019blog}.

But we have argued earlier that, in the Scully-Dr{\"u}hl-type quantum eraser as depicted in \figref{fig:SD interferometer}, the which-way information of $\gamma_s$ or $\gamma_i$ becomes factual when $\theta_1=0,\pi$ or $\theta_2=0,\pi$. Does this not contradict the MWI assertion that a quanton in a two-path interferometer \emph{always} travels through both paths?
In the MWI, whatever values of $\theta_1$ and $\theta_2$ are given, one cannot deduce the which-way information of $\gamma_s$ or $\gamma_i$ from the measurement outcome by inferential reasoning (factual or counterfactual), because the universal wavefuncton \eqref{psi f MWI} contains all the  possible outcomes $(D_+,D'_+)$, $(D_+,D'_-)$, $(D_-,D'_+)$, and $(D_-,D'_-)$, even though only one of them is obtained in a ``branched'' world. Therefore, the question whether the which-way information of $\gamma_s$ or $\gamma_i$ becomes factual is meaningless in the MWI. The Copenhagen interpretation and the MWI are incommensurable in this regard.
The only statements that are meaningful in both interpretations are the probabilities of the outcomes of $D_\pm$ and $D'_\pm$ and the correlation between them. Since the Copenhagen interpretation and the MWI yield the same probabilistic prediction of the outcomes, there is no real contradiction other than the clash of incommensurability.
But how does the MWI end up with the same prediction? Particularly, if no factual or counterfactual reasoning is employed at all, how does it yield the same prediction that the outcomes of $D_\pm$ and $D'_\pm$ must be completely correlated when $\theta_1=0,\pi$ and $\theta_2=0,\pi$? It turns out, the prediction is completely determined by the evolution, according to the MWI postulate.
In our case, the evolution is govern by \eqref{evolution gamma s} and \eqref{evolution gamma i}, which ultimately follow from the more fundamental rule \eqref{evolution path} for $\gamma_s$ and the other similar rule for $\gamma_i$. Under closer scrutiny, however, the fundamental rule \eqref{evolution path} is in fact obtained by \emph{counterfactual reasoning}! That is, \eqref{evolution gamma s} is deduced by considering what \emph{would} happen \emph{if} $\gamma_s$ travelled \emph{only} along $\Path{1}$ or $\Path{2}$. Without the counterfactual reasoning, we are unable to theorize the dynamics of evolution.
Furthermore, as the evolution dynamics depends on the parameters $\theta_1$, $\phi_1$, $\theta_2$, and $\phi_2$, which can be adjusted by Alice and Bob, intentionally or randomly at any moments, we should in principle also include Alice and Bob (or the autonomous machines that make the adjustments) in the universal wavefunction and theorize the ``dynamics'' of how they make their choices to adjust the parameters, while keeping the parameters arbitrary. This task would require us to theorize the unitary dynamics for the decision making of Alice and Bob by taking into account the molecules and chemical reactions in their brains. In practice, however, the exact dynamics of the decision making is completely irrelevant; without worrying how Alice and Bob make their choices, the evolution dynamics is theorized by the counterfactual reasoning assuming that these parameters are given \emph{definite} values.

As a demerit not often pointed out, the theoretical formalism of the MWI is not completely in harmony with its practical application: it provides an appealing ontological framework wherein the classical notion of being in a definite state is repudiated, yet in practice it still has to resort to (semi)classical reasoning about definite states in order to theorize the evolution dynamics. The necessity of invoking semiclassical reasoning is inevitable in quantum mechanics, because ``we must \emph{interpret} the experimental outcomes produced by our equipment \dots by constructing a \emph{theoretical model} of [the  macroscopic  equipment]'' (see Section 1-5 of \cite{peres1995quantum}, italics in the original).
Furthermore, the interaction between the object being measured and the measuring apparatus in principle can be ``formulated in quantum mechanical terms'' (as postulated in the MWI), ``but it must be understandable in a semiclassical way'' (see Section 12-1 of \cite{peres1995quantum}).


\section{Summary}\label{sec:summary}
The Mach-Zehnder interferometer with a nonsymmetric beam splitter $\mathrm{BS}$ as depicted in \figref{fig:interferometer} bears a close resemblance to the Stern-Gerlach experiment for spin-$1/2$ particles.
As the photon polarization states $\ket{\hat{n}_{\vartheta,\varphi},\pm}$ defined in \eqref{n + -} are analogous to the spin-up and spin-down states along the orientation $\hat{n}_{\vartheta,\varphi}=(\sin\vartheta\cos\varphi, \sin\vartheta\sin\varphi, \cos\vartheta)$, the modified Mach-Zehnder interferometer parameterized by $\theta$ and $\phi$ is exactly analogous to the Stern-Gerlach apparatus oriented in the direction $\hat{n}_{\theta,\phi}$. The probability that a signal is registered in $D_+$ or $D_-$ is given by \eqref{prob} and \eqref{inner product}, which is identical to the probability that the spin state $\ket{\hat{n}_{\vartheta,\varphi},\pm}$ entering the Stern-Gerlach apparatus oriented in the direction $\hat{n}_{\theta,\phi}$ leaves an upper or lower trace on the screen.

Correspondingly, the entanglement quantum eraser as shown in \figref{fig:two interferometers} shares exactly the same \emph{formal} structure with the EPR-Bohm experiment as shown in \figref{fig:EPR}.
The fact that the interference pattern can be recovered within each subensemble associated with $D'_+$ or $D'_-$ as shown in \eqref{prob sub} can be understood as the EPR correlation between $\ket{\hat{n}_{\theta_1,\phi_1},\pm}$ and $\ket{\hat{n}_{\theta_2,\phi_2},\pm}$ in disguise.
However, if one adopts the counterfactual but intuitive conviction that $\ket{\leftrightarrow}$ travels $\Path{1}$ and $\ket{\updownarrow}$ travels $\Path{2}$ exclusively, then the which-way information of whether $\gamma_s$ travels $\Path{1}$, $\Path{2}$, or both indeed can be affected, even retroactively, by the choice made by Bob. Unless one dismisses all counterfactual reasoning, the entanglement quantum eraser does raise the conceptual issue not found in the standard EPR puzzle, as opposed to what is claimed in \cite{kastner2019delayed,qureshi2020}.

We also demonstrate that, by the analogy \eqref{analogy}, the Scully-Dr{\"u}hl-type quantum eraser as depicted in \figref{fig:SD interferometer} is \emph{formally} equivalent to the entanglement quantum eraser as depicted in \figref{fig:two interferometers}, and yields the same quantum erasure behavior as described by \eqref{prob sub}.
Conceptually, however, the Scully-Dr{\"u}hl-type quantum eraser is more different from the EPR-Bohm experiment than the entanglement quantum eraser.
In the Scully-Dr{\"u}hl-type quantum eraser in \figref{fig:SD interferometer}, the which-way information of $\gamma_s$ is ``recorded'' in terms of the states of the two objects that are \emph{spatially separated}. The which-way information of $\gamma_s$ inferred from the measurement of the two objects becomes \emph{factual} if the measurement, such as the UQSD given by \eqref{Fs}, yields a conclusive outcome that discriminates the record.
Particularly, the which-way information of $\gamma_s$ becomes factual when $\theta_1=0,\pi$ or $\theta_2=0,\pi$, and the which-way information of $\gamma_i$ becomes factual also when $\theta_1=0,\pi$ or $\theta_2=0,\pi$.
As the which-way information of $\gamma_s$ can be ``factualized'' by Bob's choice of setting $\theta_2=0,\pi$, the fact that the which-way information of $\gamma_s$ can be influenced by the choice made by Bob is not only a consequence of counterfactual reasoning but bears some factual significance.
Therefore, it makes good sense to say that the two objects serve as the ``recorders'' of the which-way information of $\gamma_s$, and the record of the which-way information of $\gamma_s$ is ``erased'' if the potentiality to yield a conclusive outcome is eliminated by the choice made by Bob.
The Scully-Dr{\"u}hl-type quantum eraser does presents a conceptual issue beyond the standard EPR puzzle, even if counterfactual reasoning is dismissed.

We also study the nonoptimal case of quantum erasure in which the production of the signal-idler pair involves additional degrees of freedom. As shown in \eqref{prob sub nonoptimal 2}, within each subensemble associated with $D'_+$ or $D'_-$, the interference pattern can still be recovered but to a lesser extent diminished by the factor of purity $\mu_s$ compared to the optimal case.

Finally, we reconsider the quantum eraser in the MWI. The Copenhagen interpretation and the MWI make the same prediction of measurement outcomes. From the practical viewpoint, these two interpretations really makes no difference but just prescribe two different schemes of bookkeeping for calculation.
Conceptually, however, the MWI provides a very different ontological framework, wherein all possible experimental outcomes exist simultaneously in the universal wavefunction and therefore many paradoxes of quantum mechanics simply disappear as the notion of wavefunction collapse is not invoked.
In view of the MWI, the quantum eraser gives no mystery at all, since there is no such issue whether $\gamma_s$ travels along $\Path{1}$, $\Path{2}$, or both --- it unambiguously travels through both.
However, the detailed investigation of the quantum eraser reveals that the theoretical formalism of the MWI is not entirely in accord with its actual implementation.
The MWI provides an appealing ontology that completely repudiates the classical idea of being in a definite state, but in actual application it nevertheless resorts to (semi)classical reasoning about definite states in order to theorize the dynamics of evolution.


\begin{acknowledgments}
The author would like to thank Tabish Qureshi and Tai Hyun Yoon for bringing their related works to his attention and also Bo-Hung Chen and Hsiu-Chuan Hsu for having various discussions. This work was supported in part by the Ministry of Science and Technology, Taiwan under the Grant MOST 111-2112-M-110-013.
\end{acknowledgments}

\newpage


\begin{thebibliography}{46}%
\makeatletter
\providecommand \@ifxundefined [1]{%
 \@ifx{#1\undefined}
}%
\providecommand \@ifnum [1]{%
 \ifnum #1\expandafter \@firstoftwo
 \else \expandafter \@secondoftwo
 \fi
}%
\providecommand \@ifx [1]{%
 \ifx #1\expandafter \@firstoftwo
 \else \expandafter \@secondoftwo
 \fi
}%
\providecommand \natexlab [1]{#1}%
\providecommand \enquote  [1]{``#1''}%
\providecommand \bibnamefont  [1]{#1}%
\providecommand \bibfnamefont [1]{#1}%
\providecommand \citenamefont [1]{#1}%
\providecommand \href@noop [0]{\@secondoftwo}%
\providecommand \href [0]{\begingroup \@sanitize@url \@href}%
\providecommand \@href[1]{\@@startlink{#1}\@@href}%
\providecommand \@@href[1]{\endgroup#1\@@endlink}%
\providecommand \@sanitize@url [0]{\catcode `\\12\catcode `\$12\catcode
  `\&12\catcode `\#12\catcode `\^12\catcode `\_12\catcode `\%12\relax}%
\providecommand \@@startlink[1]{}%
\providecommand \@@endlink[0]{}%
\providecommand \url  [0]{\begingroup\@sanitize@url \@url }%
\providecommand \@url [1]{\endgroup\@href {#1}{\urlprefix }}%
\providecommand \urlprefix  [0]{URL }%
\providecommand \Eprint [0]{\href }%
\providecommand \doibase [0]{https://doi.org/}%
\providecommand \selectlanguage [0]{\@gobble}%
\providecommand \bibinfo  [0]{\@secondoftwo}%
\providecommand \bibfield  [0]{\@secondoftwo}%
\providecommand \translation [1]{[#1]}%
\providecommand \BibitemOpen [0]{}%
\providecommand \bibitemStop [0]{}%
\providecommand \bibitemNoStop [0]{.\EOS\space}%
\providecommand \EOS [0]{\spacefactor3000\relax}%
\providecommand \BibitemShut  [1]{\csname bibitem#1\endcsname}%
\let\auto@bib@innerbib\@empty
\bibitem [{\citenamefont {Scully}\ and\ \citenamefont
  {Dr\"uhl}(1982)}]{PhysRevA.25.2208}%
  \BibitemOpen
  \bibfield  {author} {\bibinfo {author} {\bibfnamefont {M.~O.}\ \bibnamefont
  {Scully}}\ and\ \bibinfo {author} {\bibfnamefont {K.}~\bibnamefont
  {Dr\"uhl}},\ }\bibfield  {title} {\bibinfo {title} {Quantum eraser: A
  proposed photon correlation experiment concerning observation and ``delayed
  choice'' in quantum mechanics},\ }\href
  {https://doi.org/10.1103/PhysRevA.25.2208} {\bibfield  {journal} {\bibinfo
  {journal} {Phys. Rev. A}\ }\textbf {\bibinfo {volume} {25}},\ \bibinfo
  {pages} {2208} (\bibinfo {year} {1982})}\BibitemShut {NoStop}%
\bibitem [{\citenamefont {Kim}\ \emph {et~al.}(2000)\citenamefont {Kim},
  \citenamefont {Yu}, \citenamefont {Kulik}, \citenamefont {Shih},\ and\
  \citenamefont {Scully}}]{PhysRevLett.84.1}%
  \BibitemOpen
  \bibfield  {author} {\bibinfo {author} {\bibfnamefont {Y.-H.}\ \bibnamefont
  {Kim}}, \bibinfo {author} {\bibfnamefont {R.}~\bibnamefont {Yu}}, \bibinfo
  {author} {\bibfnamefont {S.~P.}\ \bibnamefont {Kulik}}, \bibinfo {author}
  {\bibfnamefont {Y.}~\bibnamefont {Shih}},\ and\ \bibinfo {author}
  {\bibfnamefont {M.~O.}\ \bibnamefont {Scully}},\ }\bibfield  {title}
  {\bibinfo {title} {Delayed ``choice'' quantum eraser},\ }\href
  {https://doi.org/10.1103/PhysRevLett.84.1} {\bibfield  {journal} {\bibinfo
  {journal} {Phys. Rev. Lett.}\ }\textbf {\bibinfo {volume} {84}},\ \bibinfo
  {pages} {1} (\bibinfo {year} {2000})}\BibitemShut {NoStop}%
\bibitem [{\citenamefont {Walborn}\ \emph {et~al.}(2002)\citenamefont
  {Walborn}, \citenamefont {Terra~Cunha}, \citenamefont {P\'adua},\ and\
  \citenamefont {Monken}}]{PhysRevA.65.033818}%
  \BibitemOpen
  \bibfield  {author} {\bibinfo {author} {\bibfnamefont {S.~P.}\ \bibnamefont
  {Walborn}}, \bibinfo {author} {\bibfnamefont {M.~O.}\ \bibnamefont
  {Terra~Cunha}}, \bibinfo {author} {\bibfnamefont {S.}~\bibnamefont
  {P\'adua}},\ and\ \bibinfo {author} {\bibfnamefont {C.~H.}\ \bibnamefont
  {Monken}},\ }\bibfield  {title} {\bibinfo {title} {Double-slit quantum
  eraser},\ }\href {https://doi.org/10.1103/PhysRevA.65.033818} {\bibfield
  {journal} {\bibinfo  {journal} {Phys. Rev. A}\ }\textbf {\bibinfo {volume}
  {65}},\ \bibinfo {pages} {033818} (\bibinfo {year} {2002})}\BibitemShut
  {NoStop}%
\bibitem [{\citenamefont {Ma}\ \emph {et~al.}(2016)\citenamefont {Ma},
  \citenamefont {Kofler},\ and\ \citenamefont
  {Zeilinger}}]{RevModPhys.88.015005}%
  \BibitemOpen
  \bibfield  {author} {\bibinfo {author} {\bibfnamefont {X.-s.}\ \bibnamefont
  {Ma}}, \bibinfo {author} {\bibfnamefont {J.}~\bibnamefont {Kofler}},\ and\
  \bibinfo {author} {\bibfnamefont {A.}~\bibnamefont {Zeilinger}},\ }\bibfield
  {title} {\bibinfo {title} {Delayed-choice gedanken experiments and their
  realizations},\ }\href {https://doi.org/10.1103/RevModPhys.88.015005}
  {\bibfield  {journal} {\bibinfo  {journal} {Rev. Mod. Phys.}\ }\textbf
  {\bibinfo {volume} {88}},\ \bibinfo {pages} {015005} (\bibinfo {year}
  {2016})}\BibitemShut {NoStop}%
\bibitem [{\citenamefont {Chiou}\ and\ \citenamefont {Hsu}(2022)}]{chiou2022}%
  \BibitemOpen
  \bibfield  {author} {\bibinfo {author} {\bibfnamefont {D.-W.}\ \bibnamefont
  {Chiou}}\ and\ \bibinfo {author} {\bibfnamefont {H.-C.}\ \bibnamefont
  {Hsu}},\ }\href@noop {} {\bibinfo {title} {Complementarity relations of a
  delayed-choice quantum eraser in a quantum circuit}},\ \bibinfo
  {howpublished} {\url{https://arxiv.org/abs/2207.03946}} (\bibinfo {year}
  {2022})\BibitemShut {NoStop}%
\bibitem [{\citenamefont {Englert}\ \emph {et~al.}(1999)\citenamefont
  {Englert}, \citenamefont {Scully},\ and\ \citenamefont
  {Walther}}]{englert1999quantum}%
  \BibitemOpen
  \bibfield  {author} {\bibinfo {author} {\bibfnamefont {B.-G.}\ \bibnamefont
  {Englert}}, \bibinfo {author} {\bibfnamefont {M.~O.}\ \bibnamefont
  {Scully}},\ and\ \bibinfo {author} {\bibfnamefont {H.}~\bibnamefont
  {Walther}},\ }\bibfield  {title} {\bibinfo {title} {Quantum erasure in
  double-slit interferometers with which-way detectors},\ }\href
  {https://doi.org/10.1119/1.19257} {\bibfield  {journal} {\bibinfo  {journal}
  {American Journal of Physics}\ }\textbf {\bibinfo {volume} {67}},\ \bibinfo
  {pages} {325} (\bibinfo {year} {1999})}\BibitemShut {NoStop}%
\bibitem [{\citenamefont {Mohrhoff}(1999)}]{mohrhoff1999objectivity}%
  \BibitemOpen
  \bibfield  {author} {\bibinfo {author} {\bibfnamefont {U.}~\bibnamefont
  {Mohrhoff}},\ }\bibfield  {title} {\bibinfo {title} {Objectivity,
  retrocausation, and the experiment of {Englert, Scully, and Walther}},\
  }\href {https://doi.org/10.1119/1.19258} {\bibfield  {journal} {\bibinfo
  {journal} {American Journal of Physics}\ }\textbf {\bibinfo {volume} {67}},\
  \bibinfo {pages} {330} (\bibinfo {year} {1999})}\BibitemShut {NoStop}%
\bibitem [{\citenamefont {Aharonov}\ and\ \citenamefont
  {Zubairy}(2005)}]{aharonov2005time}%
  \BibitemOpen
  \bibfield  {author} {\bibinfo {author} {\bibfnamefont {Y.}~\bibnamefont
  {Aharonov}}\ and\ \bibinfo {author} {\bibfnamefont {M.~S.}\ \bibnamefont
  {Zubairy}},\ }\bibfield  {title} {\bibinfo {title} {Time and the quantum:
  Erasing the past and impacting the future},\ }\href
  {https://doi.org/10.1126/science.1107787} {\bibfield  {journal} {\bibinfo
  {journal} {Science}\ }\textbf {\bibinfo {volume} {307}},\ \bibinfo {pages}
  {875} (\bibinfo {year} {2005})}\BibitemShut {NoStop}%
\bibitem [{\citenamefont {Hiley}\ and\ \citenamefont
  {Callaghan}(2006)}]{hiley2006erased}%
  \BibitemOpen
  \bibfield  {author} {\bibinfo {author} {\bibfnamefont {B.}~\bibnamefont
  {Hiley}}\ and\ \bibinfo {author} {\bibfnamefont {R.}~\bibnamefont
  {Callaghan}},\ }\bibfield  {title} {\bibinfo {title} {What is erased in the
  quantum erasure?},\ }\href {https://doi.org/10.1007/s10701-006-9086-4}
  {\bibfield  {journal} {\bibinfo  {journal} {Foundations of Physics}\ }\textbf
  {\bibinfo {volume} {36}},\ \bibinfo {pages} {1869} (\bibinfo {year}
  {2006})}\BibitemShut {NoStop}%
\bibitem [{\citenamefont {Ellerman}(2015)}]{ellerman2015delayed}%
  \BibitemOpen
  \bibfield  {author} {\bibinfo {author} {\bibfnamefont {D.}~\bibnamefont
  {Ellerman}},\ }\bibfield  {title} {\bibinfo {title} {Why delayed choice
  experiments do not imply retrocausality},\ }\href
  {https://doi.org/10.1007/s40509-014-0026-2} {\bibfield  {journal} {\bibinfo
  {journal} {Quantum Studies: Mathematics and Foundations}\ }\textbf {\bibinfo
  {volume} {2}},\ \bibinfo {pages} {183} (\bibinfo {year} {2015})}\BibitemShut
  {NoStop}%
\bibitem [{\citenamefont {Fankhauser}(2019)}]{fankhauser2017taming}%
  \BibitemOpen
  \bibfield  {author} {\bibinfo {author} {\bibfnamefont {J.}~\bibnamefont
  {Fankhauser}},\ }\bibfield  {title} {\bibinfo {title} {Taming the delayed
  choice quantum eraser},\ }\href {https://doi.org/10.12743/quanta.v8i1.88}
  {\bibfield  {journal} {\bibinfo  {journal} {Quanta}\ }\textbf {\bibinfo
  {volume} {8}},\ \bibinfo {pages} {44} (\bibinfo {year} {2019})}\BibitemShut
  {NoStop}%
\bibitem [{\citenamefont {Kastner}(2019)}]{kastner2019delayed}%
  \BibitemOpen
  \bibfield  {author} {\bibinfo {author} {\bibfnamefont {R.}~\bibnamefont
  {Kastner}},\ }\bibfield  {title} {\bibinfo {title} {The `delayed choice
  quantum eraser' neither erases nor delays},\ }\href
  {https://doi.org/10.1007/s10701-019-00278-8} {\bibfield  {journal} {\bibinfo
  {journal} {Foundations of Physics}\ }\textbf {\bibinfo {volume} {49}},\
  \bibinfo {pages} {717} (\bibinfo {year} {2019})}\BibitemShut {NoStop}%
\bibitem [{\citenamefont {Qureshi}(2020)}]{qureshi2020}%
  \BibitemOpen
  \bibfield  {author} {\bibinfo {author} {\bibfnamefont {T.}~\bibnamefont
  {Qureshi}},\ }\bibfield  {title} {\bibinfo {title} {Demystifying the
  delayed-choice quantum eraser},\ }\href
  {https://doi.org/10.1088/1361-6404/ab923e} {\bibfield  {journal} {\bibinfo
  {journal} {European Journal of Physics}\ }\textbf {\bibinfo {volume} {41}},\
  \bibinfo {pages} {055403} (\bibinfo {year} {2020})}\BibitemShut {NoStop}%
\bibitem [{\citenamefont {Greene}(2004)}]{greene2004fabric}%
  \BibitemOpen
  \bibfield  {author} {\bibinfo {author} {\bibfnamefont {B.}~\bibnamefont
  {Greene}},\ }\href@noop {} {\emph {\bibinfo {title} {The Fabric of the
  Cosmos: Space, Time, and the Texture of Reality}}}\ (\bibinfo  {publisher}
  {Knopf},\ \bibinfo {address} {New York},\ \bibinfo {year} {2004})\BibitemShut
  {NoStop}%
\bibitem [{\citenamefont {Carrol}(2019)}]{carroll2019blog}%
  \BibitemOpen
  \bibfield  {author} {\bibinfo {author} {\bibfnamefont {S.}~\bibnamefont
  {Carrol}},\ }\href@noop {} {\bibinfo {title} {The notorious delayed-choice
  quantum eraser}},\ \bibinfo {howpublished}
  {\url{https://www.preposterousuniverse.com/blog/2019/09/21/the-notorious-delayed-choice-quantum-eraser/}}
  (\bibinfo {year} {2019})\BibitemShut {NoStop}%
\bibitem [{\citenamefont {Bohm}\ and\ \citenamefont
  {Aharonov}(1957)}]{PhysRev.108.1070}%
  \BibitemOpen
  \bibfield  {author} {\bibinfo {author} {\bibfnamefont {D.}~\bibnamefont
  {Bohm}}\ and\ \bibinfo {author} {\bibfnamefont {Y.}~\bibnamefont
  {Aharonov}},\ }\bibfield  {title} {\bibinfo {title} {Discussion of
  experimental proof for the paradox of {Einstein, Rosen, and Podolsky}},\
  }\href {https://doi.org/10.1103/PhysRev.108.1070} {\bibfield  {journal}
  {\bibinfo  {journal} {Phys. Rev.}\ }\textbf {\bibinfo {volume} {108}},\
  \bibinfo {pages} {1070} (\bibinfo {year} {1957})}\BibitemShut {NoStop}%
\bibitem [{\citenamefont {Reid}\ \emph {et~al.}(2009)\citenamefont {Reid},
  \citenamefont {Drummond}, \citenamefont {Bowen}, \citenamefont {Cavalcanti},
  \citenamefont {Lam}, \citenamefont {Bachor}, \citenamefont {Andersen},\ and\
  \citenamefont {Leuchs}}]{RevModPhys.81.1727}%
  \BibitemOpen
  \bibfield  {author} {\bibinfo {author} {\bibfnamefont {M.~D.}\ \bibnamefont
  {Reid}}, \bibinfo {author} {\bibfnamefont {P.~D.}\ \bibnamefont {Drummond}},
  \bibinfo {author} {\bibfnamefont {W.~P.}\ \bibnamefont {Bowen}}, \bibinfo
  {author} {\bibfnamefont {E.~G.}\ \bibnamefont {Cavalcanti}}, \bibinfo
  {author} {\bibfnamefont {P.~K.}\ \bibnamefont {Lam}}, \bibinfo {author}
  {\bibfnamefont {H.~A.}\ \bibnamefont {Bachor}}, \bibinfo {author}
  {\bibfnamefont {U.~L.}\ \bibnamefont {Andersen}},\ and\ \bibinfo {author}
  {\bibfnamefont {G.}~\bibnamefont {Leuchs}},\ }\bibfield  {title} {\bibinfo
  {title} {Colloquium: The {Einstein-Podolsky-Rosen} paradox: From concepts to
  applications},\ }\href {https://doi.org/10.1103/RevModPhys.81.1727}
  {\bibfield  {journal} {\bibinfo  {journal} {Rev. Mod. Phys.}\ }\textbf
  {\bibinfo {volume} {81}},\ \bibinfo {pages} {1727} (\bibinfo {year}
  {2009})}\BibitemShut {NoStop}%
\bibitem [{\citenamefont {Everett}(1957)}]{RevModPhys.29.454}%
  \BibitemOpen
  \bibfield  {author} {\bibinfo {author} {\bibfnamefont {H.}~\bibnamefont
  {Everett}},\ }\bibfield  {title} {\bibinfo {title} {``{Relative} state''
  formulation of quantum mechanics},\ }\href
  {https://doi.org/10.1103/RevModPhys.29.454} {\bibfield  {journal} {\bibinfo
  {journal} {Rev. Mod. Phys.}\ }\textbf {\bibinfo {volume} {29}},\ \bibinfo
  {pages} {454} (\bibinfo {year} {1957})}\BibitemShut {NoStop}%
\bibitem [{\citenamefont {DeWitt}\ and\ \citenamefont
  {Graham}(1975)}]{dewitt1975many}%
  \BibitemOpen
  \bibinfo {editor} {\bibfnamefont {B.~S.}\ \bibnamefont {DeWitt}}\ and\
  \bibinfo {editor} {\bibfnamefont {N.}~\bibnamefont {Graham}},\ eds.,\
  \href@noop {} {\emph {\bibinfo {title} {The Many-Worlds Interpretation of
  Quantum Mechanics}}}\ (\bibinfo  {publisher} {Princeton University Press},\
  \bibinfo {address} {Princeton, New Jersey},\ \bibinfo {year}
  {1975})\BibitemShut {NoStop}%
\bibitem [{\citenamefont {Wheeler}(1957)}]{RevModPhys.29.463}%
  \BibitemOpen
  \bibfield  {author} {\bibinfo {author} {\bibfnamefont {J.~A.}\ \bibnamefont
  {Wheeler}},\ }\bibfield  {title} {\bibinfo {title} {Assessment of {Everett's}
  ``relative state'' formulation of quantum theory},\ }\href
  {https://doi.org/10.1103/RevModPhys.29.463} {\bibfield  {journal} {\bibinfo
  {journal} {Rev. Mod. Phys.}\ }\textbf {\bibinfo {volume} {29}},\ \bibinfo
  {pages} {463} (\bibinfo {year} {1957})}\BibitemShut {NoStop}%
\bibitem [{\citenamefont {DeWitt}(1970)}]{dewitt1970}%
  \BibitemOpen
  \bibfield  {author} {\bibinfo {author} {\bibfnamefont {B.~S.}\ \bibnamefont
  {DeWitt}},\ }\bibfield  {title} {\bibinfo {title} {Quantum mechanics and
  reality},\ }\href {https://doi.org/10.1063/1.3022331} {\bibfield  {journal}
  {\bibinfo  {journal} {Physics Today}\ }\textbf {\bibinfo {volume} {23}},\
  \bibinfo {pages} {30} (\bibinfo {year} {1970})}\BibitemShut {NoStop}%
\bibitem [{\citenamefont {Peres}(1995)}]{peres1995quantum}%
  \BibitemOpen
  \bibfield  {author} {\bibinfo {author} {\bibfnamefont {A.}~\bibnamefont
  {Peres}},\ }\href@noop {} {\emph {\bibinfo {title} {Quantum Theory: Methods
  and Concepts}}}\ (\bibinfo  {publisher} {Kluwer Academic Publishers},\
  \bibinfo {address} {Dordrecht, Netherlands},\ \bibinfo {year}
  {1995})\BibitemShut {NoStop}%
\bibitem [{\citenamefont {Chiou}(2013)}]{Chiou_2013}%
  \BibitemOpen
  \bibfield  {author} {\bibinfo {author} {\bibfnamefont {D.-W.}\ \bibnamefont
  {Chiou}},\ }\bibfield  {title} {\bibinfo {title} {Timeless path integral for
  relativistic quantum mechanics},\ }\href
  {https://doi.org/10.1088/0264-9381/30/12/125004} {\bibfield  {journal}
  {\bibinfo  {journal} {Classical and Quantum Gravity}\ }\textbf {\bibinfo
  {volume} {30}},\ \bibinfo {pages} {125004} (\bibinfo {year}
  {2013})}\BibitemShut {NoStop}%
\bibitem [{\citenamefont {Wheeler}(1983)}]{wheeler1983}%
  \BibitemOpen
  \bibfield  {author} {\bibinfo {author} {\bibfnamefont {J.~A.}\ \bibnamefont
  {Wheeler}},\ }\bibinfo {title} {Law without law},\ in\ \href@noop {} {\emph
  {\bibinfo {booktitle} {Quantum Theory and Measurement}}},\ \bibinfo {editor}
  {edited by\ \bibinfo {editor} {\bibfnamefont {J.~A.}\ \bibnamefont
  {Wheeler}}\ and\ \bibinfo {editor} {\bibfnamefont {W.~H.}\ \bibnamefont
  {Zurek}}}\ (\bibinfo  {publisher} {Princeton University Press},\ \bibinfo
  {address} {Princeton, New Jersey},\ \bibinfo {year} {1983})\ pp.\ \bibinfo
  {pages} {182--213}\BibitemShut {NoStop}%
\bibitem [{\citenamefont {Alley}\ \emph {et~al.}(1986)\citenamefont {Alley},
  \citenamefont {Jakubowicz},\ and\ \citenamefont {Wickes}}]{alley1986delayed}%
  \BibitemOpen
  \bibfield  {author} {\bibinfo {author} {\bibfnamefont {C.~O.}\ \bibnamefont
  {Alley}}, \bibinfo {author} {\bibfnamefont {O.~G.}\ \bibnamefont
  {Jakubowicz}},\ and\ \bibinfo {author} {\bibfnamefont {W.~C.}\ \bibnamefont
  {Wickes}},\ }\bibfield  {title} {\bibinfo {title} {Results of the
  delayed-random-choice quantum mechanics experiment with light quanta},\ }in\
  \href@noop {} {\emph {\bibinfo {booktitle} {Proceedings of the 2nd
  International Symposium on Foundations of Quantum Mechanics}}}\ (\bibinfo
  {address} {Tokyo},\ \bibinfo {year} {1986})\ p.~\bibinfo {pages}
  {36}\BibitemShut {NoStop}%
\bibitem [{\citenamefont {Hellmuth}\ \emph {et~al.}(1987)\citenamefont
  {Hellmuth}, \citenamefont {Walther}, \citenamefont {Zajonc},\ and\
  \citenamefont {Schleich}}]{PhysRevA.35.2532}%
  \BibitemOpen
  \bibfield  {author} {\bibinfo {author} {\bibfnamefont {T.}~\bibnamefont
  {Hellmuth}}, \bibinfo {author} {\bibfnamefont {H.}~\bibnamefont {Walther}},
  \bibinfo {author} {\bibfnamefont {A.}~\bibnamefont {Zajonc}},\ and\ \bibinfo
  {author} {\bibfnamefont {W.}~\bibnamefont {Schleich}},\ }\bibfield  {title}
  {\bibinfo {title} {Delayed-choice experiments in quantum interference},\
  }\href {https://doi.org/10.1103/PhysRevA.35.2532} {\bibfield  {journal}
  {\bibinfo  {journal} {Phys. Rev. A}\ }\textbf {\bibinfo {volume} {35}},\
  \bibinfo {pages} {2532} (\bibinfo {year} {1987})}\BibitemShut {NoStop}%
\bibitem [{\citenamefont {Baldzuhn}\ \emph {et~al.}(1989)\citenamefont
  {Baldzuhn}, \citenamefont {Mohler},\ and\ \citenamefont
  {Martienssen}}]{baldzuhn1989wave}%
  \BibitemOpen
  \bibfield  {author} {\bibinfo {author} {\bibfnamefont {J.}~\bibnamefont
  {Baldzuhn}}, \bibinfo {author} {\bibfnamefont {E.}~\bibnamefont {Mohler}},\
  and\ \bibinfo {author} {\bibfnamefont {W.}~\bibnamefont {Martienssen}},\
  }\bibfield  {title} {\bibinfo {title} {A wave-particle delayed-choice
  experiment with a single-photon state},\ }\href
  {https://doi.org/10.1007/BF01313681} {\bibfield  {journal} {\bibinfo
  {journal} {Zeitschrift f{\"u}r Physik B Condensed Matter}\ }\textbf {\bibinfo
  {volume} {77}},\ \bibinfo {pages} {347} (\bibinfo {year} {1989})}\BibitemShut
  {NoStop}%
\bibitem [{\citenamefont {Lawson-Daku}\ \emph {et~al.}(1996)\citenamefont
  {Lawson-Daku}, \citenamefont {Asimov}, \citenamefont {Gorceix}, \citenamefont
  {Miniatura}, \citenamefont {Robert},\ and\ \citenamefont
  {Baudon}}]{PhysRevA.54.5042}%
  \BibitemOpen
  \bibfield  {author} {\bibinfo {author} {\bibfnamefont {B.~J.}\ \bibnamefont
  {Lawson-Daku}}, \bibinfo {author} {\bibfnamefont {R.}~\bibnamefont {Asimov}},
  \bibinfo {author} {\bibfnamefont {O.}~\bibnamefont {Gorceix}}, \bibinfo
  {author} {\bibfnamefont {C.}~\bibnamefont {Miniatura}}, \bibinfo {author}
  {\bibfnamefont {J.}~\bibnamefont {Robert}},\ and\ \bibinfo {author}
  {\bibfnamefont {J.}~\bibnamefont {Baudon}},\ }\bibfield  {title} {\bibinfo
  {title} {Delayed choices in atom {Stern-Gerlach} interferometry},\ }\href
  {https://doi.org/10.1103/PhysRevA.54.5042} {\bibfield  {journal} {\bibinfo
  {journal} {Phys. Rev. A}\ }\textbf {\bibinfo {volume} {54}},\ \bibinfo
  {pages} {5042} (\bibinfo {year} {1996})}\BibitemShut {NoStop}%
\bibitem [{\citenamefont {Kawai}\ \emph {et~al.}(1998)\citenamefont {Kawai},
  \citenamefont {Ebisawa}, \citenamefont {Tasaki}, \citenamefont {Hino},
  \citenamefont {Yamazaki}, \citenamefont {Akiyoshi}, \citenamefont
  {Matsumoto}, \citenamefont {Achiwa},\ and\ \citenamefont
  {Otake}}]{KAWAI1998259}%
  \BibitemOpen
  \bibfield  {author} {\bibinfo {author} {\bibfnamefont {T.}~\bibnamefont
  {Kawai}}, \bibinfo {author} {\bibfnamefont {T.}~\bibnamefont {Ebisawa}},
  \bibinfo {author} {\bibfnamefont {S.}~\bibnamefont {Tasaki}}, \bibinfo
  {author} {\bibfnamefont {M.}~\bibnamefont {Hino}}, \bibinfo {author}
  {\bibfnamefont {D.}~\bibnamefont {Yamazaki}}, \bibinfo {author}
  {\bibfnamefont {T.}~\bibnamefont {Akiyoshi}}, \bibinfo {author}
  {\bibfnamefont {Y.}~\bibnamefont {Matsumoto}}, \bibinfo {author}
  {\bibfnamefont {N.}~\bibnamefont {Achiwa}},\ and\ \bibinfo {author}
  {\bibfnamefont {Y.}~\bibnamefont {Otake}},\ }\bibfield  {title} {\bibinfo
  {title} {Realization of a delayed choice experiment using a multilayer cold
  neutron pulser},\ }\href {https://doi.org/10.1016/S0168-9002(98)00263-0}
  {\bibfield  {journal} {\bibinfo  {journal} {Nuclear Instruments and Methods
  in Physics Research Section A: Accelerators, Spectrometers, Detectors and
  Associated Equipment}\ }\textbf {\bibinfo {volume} {410}},\ \bibinfo {pages}
  {259} (\bibinfo {year} {1998})}\BibitemShut {NoStop}%
\bibitem [{\citenamefont {Jacques}\ \emph {et~al.}(2007)\citenamefont
  {Jacques}, \citenamefont {Wu}, \citenamefont {Grosshans}, \citenamefont
  {Treussart}, \citenamefont {Grangier}, \citenamefont {Aspect},\ and\
  \citenamefont {Roch}}]{jacques2007experimental}%
  \BibitemOpen
  \bibfield  {author} {\bibinfo {author} {\bibfnamefont {V.}~\bibnamefont
  {Jacques}}, \bibinfo {author} {\bibfnamefont {E.}~\bibnamefont {Wu}},
  \bibinfo {author} {\bibfnamefont {F.}~\bibnamefont {Grosshans}}, \bibinfo
  {author} {\bibfnamefont {F.}~\bibnamefont {Treussart}}, \bibinfo {author}
  {\bibfnamefont {P.}~\bibnamefont {Grangier}}, \bibinfo {author}
  {\bibfnamefont {A.}~\bibnamefont {Aspect}},\ and\ \bibinfo {author}
  {\bibfnamefont {J.-F.}\ \bibnamefont {Roch}},\ }\bibfield  {title} {\bibinfo
  {title} {Experimental realization of {Wheeler's} delayed-choice gedanken
  experiment},\ }\href {https://doi.org/10.1126/science.1136303} {\bibfield
  {journal} {\bibinfo  {journal} {Science}\ }\textbf {\bibinfo {volume}
  {315}},\ \bibinfo {pages} {966} (\bibinfo {year} {2007})}\BibitemShut
  {NoStop}%
\bibitem [{\citenamefont {Marolf}\ and\ \citenamefont
  {Rovelli}(2002)}]{PhysRevD.66.023510}%
  \BibitemOpen
  \bibfield  {author} {\bibinfo {author} {\bibfnamefont {D.}~\bibnamefont
  {Marolf}}\ and\ \bibinfo {author} {\bibfnamefont {C.}~\bibnamefont
  {Rovelli}},\ }\bibfield  {title} {\bibinfo {title} {Relativistic quantum
  measurement},\ }\href {https://doi.org/10.1103/PhysRevD.66.023510} {\bibfield
   {journal} {\bibinfo  {journal} {Phys. Rev. D}\ }\textbf {\bibinfo {volume}
  {66}},\ \bibinfo {pages} {023510} (\bibinfo {year} {2002})}\BibitemShut
  {NoStop}%
\bibitem [{\citenamefont {Dowker}\ and\ \citenamefont
  {Kent}(1995)}]{PhysRevLett.75.3038}%
  \BibitemOpen
  \bibfield  {author} {\bibinfo {author} {\bibfnamefont {F.}~\bibnamefont
  {Dowker}}\ and\ \bibinfo {author} {\bibfnamefont {A.}~\bibnamefont {Kent}},\
  }\bibfield  {title} {\bibinfo {title} {Properties of consistent histories},\
  }\href {https://doi.org/10.1103/PhysRevLett.75.3038} {\bibfield  {journal}
  {\bibinfo  {journal} {Phys. Rev. Lett.}\ }\textbf {\bibinfo {volume} {75}},\
  \bibinfo {pages} {3038} (\bibinfo {year} {1995})}\BibitemShut {NoStop}%
\bibitem [{\citenamefont {Qureshi}(2021)}]{qureshi2021delayed}%
  \BibitemOpen
  \bibfield  {author} {\bibinfo {author} {\bibfnamefont {T.}~\bibnamefont
  {Qureshi}},\ }\bibfield  {title} {\bibinfo {title} {The delayed-choice
  quantum eraser leaves no choice},\ }\href
  {https://doi.org/10.1007/s10773-021-04906-w} {\bibfield  {journal} {\bibinfo
  {journal} {International Journal of Theoretical Physics}\ }\textbf {\bibinfo
  {volume} {60}},\ \bibinfo {pages} {3076} (\bibinfo {year}
  {2021})}\BibitemShut {NoStop}%
\bibitem [{\citenamefont {Qian}\ and\ \citenamefont
  {Agarwal}(2020)}]{PhysRevResearch.2.012031}%
  \BibitemOpen
  \bibfield  {author} {\bibinfo {author} {\bibfnamefont {X.-F.}\ \bibnamefont
  {Qian}}\ and\ \bibinfo {author} {\bibfnamefont {G.~S.}\ \bibnamefont
  {Agarwal}},\ }\bibfield  {title} {\bibinfo {title} {Quantum duality: A source
  point of view},\ }\href {https://doi.org/10.1103/PhysRevResearch.2.012031}
  {\bibfield  {journal} {\bibinfo  {journal} {Phys. Rev. Research}\ }\textbf
  {\bibinfo {volume} {2}},\ \bibinfo {pages} {012031} (\bibinfo {year}
  {2020})}\BibitemShut {NoStop}%
\bibitem [{\citenamefont {Anthony}(2000)}]{anthony2000quantum}%
  \BibitemOpen
  \bibfield  {author} {\bibinfo {author} {\bibfnamefont {C.}~\bibnamefont
  {Anthony}},\ }\bibfield  {title} {\bibinfo {title} {Quantum state
  discrimination},\ }\href {https://doi.org/10.1080/00107510010002599}
  {\bibfield  {journal} {\bibinfo  {journal} {Contemporary Physics}\ }\textbf
  {\bibinfo {volume} {41}},\ \bibinfo {pages} {401} (\bibinfo {year}
  {2000})}\BibitemShut {NoStop}%
\bibitem [{\citenamefont {Bergou}\ \emph {et~al.}(2004)\citenamefont {Bergou},
  \citenamefont {Herzog},\ and\ \citenamefont {Hillery}}]{bergou2004quantum}%
  \BibitemOpen
  \bibfield  {author} {\bibinfo {author} {\bibfnamefont {J.}~\bibnamefont
  {Bergou}}, \bibinfo {author} {\bibfnamefont {U.}~\bibnamefont {Herzog}},\
  and\ \bibinfo {author} {\bibfnamefont {M.}~\bibnamefont {Hillery}},\
  }\bibinfo {title} {Discrimination of quantum states},\ in\ \href
  {https://doi.org/10.1007/978-3-540-44481-7_11} {\emph {\bibinfo {booktitle}
  {Quantum State Estimation. Lecture Notes in Physics, vol 64}}}\ (\bibinfo
  {publisher} {Springer},\ \bibinfo {address} {Berlin, Heidelberg},\ \bibinfo
  {year} {2004})\ pp.\ \bibinfo {pages} {417--465}\BibitemShut {NoStop}%
\bibitem [{\citenamefont {Coles}\ \emph {et~al.}(2014)\citenamefont {Coles},
  \citenamefont {Kaniewski},\ and\ \citenamefont
  {Wehner}}]{coles2014equivalence}%
  \BibitemOpen
  \bibfield  {author} {\bibinfo {author} {\bibfnamefont {P.~J.}\ \bibnamefont
  {Coles}}, \bibinfo {author} {\bibfnamefont {J.}~\bibnamefont {Kaniewski}},\
  and\ \bibinfo {author} {\bibfnamefont {S.}~\bibnamefont {Wehner}},\
  }\bibfield  {title} {\bibinfo {title} {Equivalence of wave-particle duality
  to entropic uncertainty},\ }\href {https://doi.org/10.1038/ncomms6814}
  {\bibfield  {journal} {\bibinfo  {journal} {Nature communications}\ }\textbf
  {\bibinfo {volume} {5}},\ \bibinfo {pages} {1} (\bibinfo {year}
  {2014})}\BibitemShut {NoStop}%
\bibitem [{\citenamefont {Coles}(2016)}]{PhysRevA.93.062111}%
  \BibitemOpen
  \bibfield  {author} {\bibinfo {author} {\bibfnamefont {P.~J.}\ \bibnamefont
  {Coles}},\ }\bibfield  {title} {\bibinfo {title} {Entropic framework for
  wave-particle duality in multipath interferometers},\ }\href
  {https://doi.org/10.1103/PhysRevA.93.062111} {\bibfield  {journal} {\bibinfo
  {journal} {Phys. Rev. A}\ }\textbf {\bibinfo {volume} {93}},\ \bibinfo
  {pages} {062111} (\bibinfo {year} {2016})}\BibitemShut {NoStop}%
\bibitem [{\citenamefont {Jaeger}\ \emph {et~al.}(1995)\citenamefont {Jaeger},
  \citenamefont {Shimony},\ and\ \citenamefont {Vaidman}}]{PhysRevA.51.54}%
  \BibitemOpen
  \bibfield  {author} {\bibinfo {author} {\bibfnamefont {G.}~\bibnamefont
  {Jaeger}}, \bibinfo {author} {\bibfnamefont {A.}~\bibnamefont {Shimony}},\
  and\ \bibinfo {author} {\bibfnamefont {L.}~\bibnamefont {Vaidman}},\
  }\bibfield  {title} {\bibinfo {title} {Two interferometric
  complementarities},\ }\href {https://doi.org/10.1103/PhysRevA.51.54}
  {\bibfield  {journal} {\bibinfo  {journal} {Phys. Rev. A}\ }\textbf {\bibinfo
  {volume} {51}},\ \bibinfo {pages} {54} (\bibinfo {year} {1995})}\BibitemShut
  {NoStop}%
\bibitem [{\citenamefont {Englert}(1996)}]{PhysRevLett.77.2154}%
  \BibitemOpen
  \bibfield  {author} {\bibinfo {author} {\bibfnamefont {B.-G.}\ \bibnamefont
  {Englert}},\ }\bibfield  {title} {\bibinfo {title} {Fringe visibility and
  which-way information: An inequality},\ }\href
  {https://doi.org/10.1103/PhysRevLett.77.2154} {\bibfield  {journal} {\bibinfo
   {journal} {Phys. Rev. Lett.}\ }\textbf {\bibinfo {volume} {77}},\ \bibinfo
  {pages} {2154} (\bibinfo {year} {1996})}\BibitemShut {NoStop}%
\bibitem [{\citenamefont {Eichmann}\ \emph {et~al.}(1993)\citenamefont
  {Eichmann}, \citenamefont {Bergquist}, \citenamefont {Bollinger},
  \citenamefont {Gilligan}, \citenamefont {Itano}, \citenamefont {Wineland},\
  and\ \citenamefont {Raizen}}]{PhysRevLett.70.2359}%
  \BibitemOpen
  \bibfield  {author} {\bibinfo {author} {\bibfnamefont {U.}~\bibnamefont
  {Eichmann}}, \bibinfo {author} {\bibfnamefont {J.~C.}\ \bibnamefont
  {Bergquist}}, \bibinfo {author} {\bibfnamefont {J.~J.}\ \bibnamefont
  {Bollinger}}, \bibinfo {author} {\bibfnamefont {J.~M.}\ \bibnamefont
  {Gilligan}}, \bibinfo {author} {\bibfnamefont {W.~M.}\ \bibnamefont {Itano}},
  \bibinfo {author} {\bibfnamefont {D.~J.}\ \bibnamefont {Wineland}},\ and\
  \bibinfo {author} {\bibfnamefont {M.~G.}\ \bibnamefont {Raizen}},\ }\bibfield
   {title} {\bibinfo {title} {Young's interference experiment with light
  scattered from two atoms},\ }\href
  {https://doi.org/10.1103/PhysRevLett.70.2359} {\bibfield  {journal} {\bibinfo
   {journal} {Phys. Rev. Lett.}\ }\textbf {\bibinfo {volume} {70}},\ \bibinfo
  {pages} {2359} (\bibinfo {year} {1993})}\BibitemShut {NoStop}%
\bibitem [{\citenamefont {Itano}\ \emph {et~al.}(1998)\citenamefont {Itano},
  \citenamefont {Bergquist}, \citenamefont {Bollinger}, \citenamefont
  {Wineland}, \citenamefont {Eichmann},\ and\ \citenamefont
  {Raizen}}]{PhysRevA.57.4176}%
  \BibitemOpen
  \bibfield  {author} {\bibinfo {author} {\bibfnamefont {W.~M.}\ \bibnamefont
  {Itano}}, \bibinfo {author} {\bibfnamefont {J.~C.}\ \bibnamefont
  {Bergquist}}, \bibinfo {author} {\bibfnamefont {J.~J.}\ \bibnamefont
  {Bollinger}}, \bibinfo {author} {\bibfnamefont {D.~J.}\ \bibnamefont
  {Wineland}}, \bibinfo {author} {\bibfnamefont {U.}~\bibnamefont {Eichmann}},\
  and\ \bibinfo {author} {\bibfnamefont {M.~G.}\ \bibnamefont {Raizen}},\
  }\bibfield  {title} {\bibinfo {title} {Complementarity and {Young's}
  interference fringes from two atoms},\ }\href
  {https://doi.org/10.1103/PhysRevA.57.4176} {\bibfield  {journal} {\bibinfo
  {journal} {Phys. Rev. A}\ }\textbf {\bibinfo {volume} {57}},\ \bibinfo
  {pages} {4176} (\bibinfo {year} {1998})}\BibitemShut {NoStop}%
\bibitem [{\citenamefont {Araneda}\ \emph {et~al.}(2018)\citenamefont
  {Araneda}, \citenamefont {Higginbottom}, \citenamefont
  {Slodi\ifmmode~\check{c}\else \v{c}\fi{}ka}, \citenamefont {Colombe},\ and\
  \citenamefont {Blatt}}]{PhysRevLett.120.193603}%
  \BibitemOpen
  \bibfield  {author} {\bibinfo {author} {\bibfnamefont {G.}~\bibnamefont
  {Araneda}}, \bibinfo {author} {\bibfnamefont {D.~B.}\ \bibnamefont
  {Higginbottom}}, \bibinfo {author} {\bibfnamefont {L.}~\bibnamefont
  {Slodi\ifmmode~\check{c}\else \v{c}\fi{}ka}}, \bibinfo {author}
  {\bibfnamefont {Y.}~\bibnamefont {Colombe}},\ and\ \bibinfo {author}
  {\bibfnamefont {R.}~\bibnamefont {Blatt}},\ }\bibfield  {title} {\bibinfo
  {title} {Interference of single photons emitted by entangled atoms in free
  space},\ }\href {https://doi.org/10.1103/PhysRevLett.120.193603} {\bibfield
  {journal} {\bibinfo  {journal} {Phys. Rev. Lett.}\ }\textbf {\bibinfo
  {volume} {120}},\ \bibinfo {pages} {193603} (\bibinfo {year}
  {2018})}\BibitemShut {NoStop}%
\bibitem [{\citenamefont {Kyung}\ \emph {et~al.}(2018)\citenamefont {Kyung},
  \citenamefont {Han}, \citenamefont {Yoon},\ and\ \citenamefont
  {Minhaeng}}]{kyung2018frequency}%
  \BibitemOpen
  \bibfield  {author} {\bibinfo {author} {\bibfnamefont {L.~S.}\ \bibnamefont
  {Kyung}}, \bibinfo {author} {\bibfnamefont {N.~S.}\ \bibnamefont {Han}},
  \bibinfo {author} {\bibfnamefont {T.~H.}\ \bibnamefont {Yoon}},\ and\
  \bibinfo {author} {\bibfnamefont {C.}~\bibnamefont {Minhaeng}},\ }\bibfield
  {title} {\bibinfo {title} {Frequency comb single-photon interferometry},\
  }\href {https://doi.org/10.1038/s42005-018-0051-2} {\bibfield  {journal}
  {\bibinfo  {journal} {Communications Physics}\ }\textbf {\bibinfo {volume}
  {1}},\ \bibinfo {pages} {51} (\bibinfo {year} {2018})}\BibitemShut {NoStop}%
\bibitem [{\citenamefont {Yoon}\ and\ \citenamefont
  {Cho}(2021)}]{sciadv.abi9268}%
  \BibitemOpen
  \bibfield  {author} {\bibinfo {author} {\bibfnamefont {T.~H.}\ \bibnamefont
  {Yoon}}\ and\ \bibinfo {author} {\bibfnamefont {M.}~\bibnamefont {Cho}},\
  }\bibfield  {title} {\bibinfo {title} {Quantitative complementarity of
  wave-particle duality},\ }\href {https://doi.org/10.1126/sciadv.abi9268}
  {\bibfield  {journal} {\bibinfo  {journal} {Science Advances}\ }\textbf
  {\bibinfo {volume} {7}},\ \bibinfo {pages} {eabi9268} (\bibinfo {year}
  {2021})}\BibitemShut {NoStop}%
\bibitem [{\citenamefont {Tegmark}(1998)}]{tegmark1998}%
  \BibitemOpen
  \bibfield  {author} {\bibinfo {author} {\bibfnamefont {M.}~\bibnamefont
  {Tegmark}},\ }\bibfield  {title} {\bibinfo {title} {The interpretation of
  quantum mechanics: Many worlds or many words?},\ }\href
  {https://doi.org/10.1002/(SICI)1521-3978(199811)46:6/8<855::AID-PROP855>3.0.CO;2-Q}
  {\bibfield  {journal} {\bibinfo  {journal} {Fortschritte der Physik}\
  }\textbf {\bibinfo {volume} {46}},\ \bibinfo {pages} {855} (\bibinfo {year}
  {1998})}\BibitemShut {NoStop}%
\end{thebibliography}
%

%

\end{document}